\documentclass[aps,preprintnumbers,showpacs,nofootinbib]{revtex4}
\usepackage[english]{babel}
\usepackage{graphicx}% Include figure files
\usepackage{dcolumn}% Align table columns on decimal point
\usepackage{bm}% bold math
\usepackage{amssymb}
\usepackage{amsmath}
\usepackage{amsfonts}
\usepackage{subfig}
\begin{document}
\newcommand{\newc}{\newcommand}
\newc{\ra}{\rightarrow}
\newc{\lra}{\leftrightarrow}
\newc{\beq}{\begin{equation}}
\newc{\eeq}{\end{equation}}
\newc{\barr}{\begin{eqnarray}}
\newc{\earr}{\end{eqnarray}}

\title{Neutral Current Coherent Cross Sections- Implications on\\
Gaseous Spherical TPC's  for detecting SN and Earth neutrinos} 
\author{J.~D.~Vergados$^{1*}$ and Y.~ Giomataris$^{2} $}
% \author{Y.~ Giomataris$^{} $}
\affiliation{ ARC Centre of Excellence in Particle Physics at the Terascale and Centre for the \\Subatomic Structure of Matter (CSSM), University of Adelaide, Adelaide SA 5005, Australia,
\footnote{Permanent address:Theoretical Physics,University of Ioannina, Ioannina, Gr 451 10, Greece.},
$^2$ IRFU, CEA, Universit\'e Paris-Saclay, F-91191 Gif-sur-Yvette, France }
\begin{abstract}
The detection of galactic supernova (SN) neutrinos represents one of the future frontiers of low-energy neutrino physics and astrophysics. The neutron coherence of neutral currents (NC) allows quite large cross sections in the case  of neutron rich targets, which can be exploited in detecting  earth and sky neutrinos by measuring nuclear recoils. They are relatively cheap and easy to maintain.
 %The collapse of a neutron star liberates a gravitational binding energy of about $3\times10^{53}$ erg, 99$\%$ of
%which is transferred to neutrinos and antineutrions  of all the flavors and only 1$\%$ to the kinetic energy of the explosion. In other words,
%a core-collapse supernova represents one of the most powerful source of (anti)neutrinos in the Universe.  
These (NC) cross sections are not dependent on flavor conversions and, thus, their measurement will provide useful information about the neutrino source. In particular they will yield information about the primary neutrino fluxes and perhaps about the spectrum after  flavor conversions in neutrino sphere.They might also provide some clues about the neutrino mass hierarchy. The advantages of large gaseous low threshold and high resolution time projection counters (TPC) detectors  TPC detectors are  discussed.

\end{abstract}


%\date{November 28, 2010}

\maketitle
\section{Introduction}
The detection of galactic supernova (SN) neutrinos represents one of the future frontiers of low-energy neutrino physics and astrophysics. In this paper we are going to discuss the relevant physics for the design and construction of a gaseous spherical TPC  for dedicated supernova detection, exploiting the coherent neutrino-nucleus elastic scattering due to the neutral current interaction. This detector can draw on the progress made in recent years in connection with measuring nuclear recoils in dark matter searches. It has low threshold and high resolution, it is relatively cheap and easy to maintain. Before doing this, however, we will briefly discuss the essential physics of neutrinos emitted in supernova explosions \cite{RAFFELT04} (for a review, see, e.g. the recent report \cite{JLMMM06})
\section{The supernova neutrino flux.}
We will assume that the neutrino spectrum can be described be a Fermi Dirac Distribution with a given temperature $T$ and a chemical potential $\mu=a T$. The constants $T$ and   $a$ will be treated as free parameters. Thus
\beq
f_{sp}(E_{\nu},T,a)={\cal N}\frac{1}{1+\exp{(E_{\nu}/T-a)}}
\eeq
where $\cal{N}$ is a normalization constant. The temperature $T$ is taken to be $3.5$, $5$ and $8$ MeV for electron neutrinos ($\nu_e$), electron antineutrinos ($\tilde{\nu}_e$) and all other flavors ($\nu_x$)  respectively. The parameter $a$ will be taken to be $0\le a \le 5$.
     \begin{figure}[!ht]
 \begin{center}
 \subfloat[]
{
\rotatebox{90}{\hspace{-0.0cm}{$f_{sp}\rightarrow$MeV$^{-1}$}}
%\rotatebox{90}{\hspace{-0.0cm} {$\Phi(x)\longrightarrow 4 \pi G_Na^2 \rho_0$}}
\includegraphics[scale=0.7]{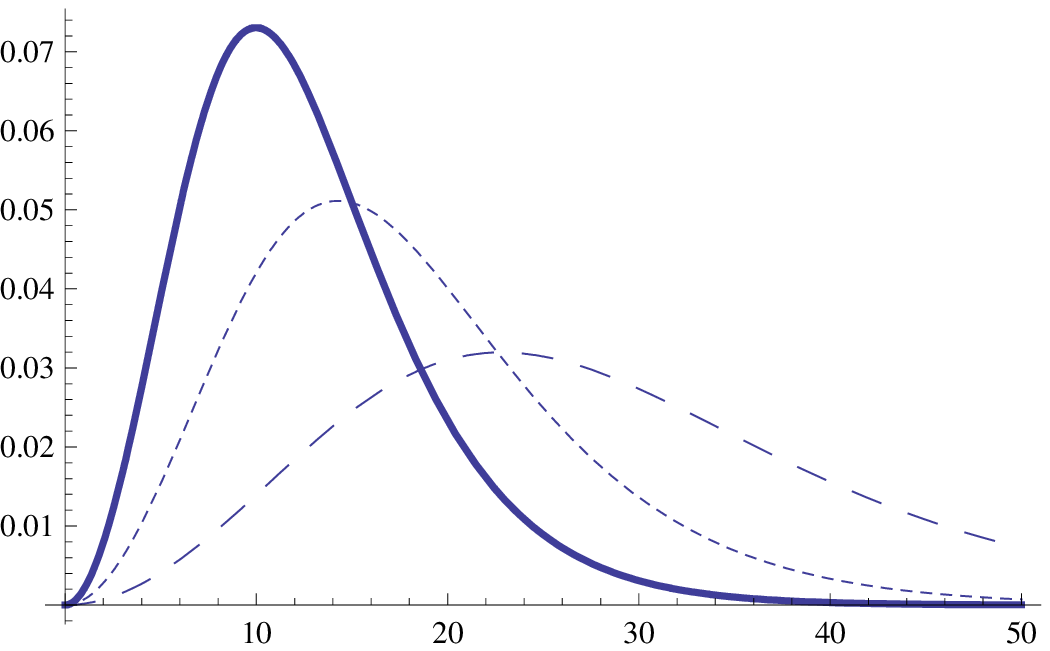}
}\\
%{\hspace{-2.0cm} {$\frac{\Phi(x)}{\Phi_0}\longrightarrow $}}
 \subfloat[]
 {
%\rotatebox{90}{\hspace{-0.0cm}{$f_{sp}\rightarrow$MeV$^{-1}$}}
%\rotatebox{90}{\hspace{-0.0cm} {$\Phi(x)\longrightarrow 4 \pi G_Na^2 \rho_0$}}
\rotatebox{90}{\hspace{-0.0cm}{$f_{sp}\rightarrow$MeV$^{-1}$ NH}}
\includegraphics[scale=0.6]{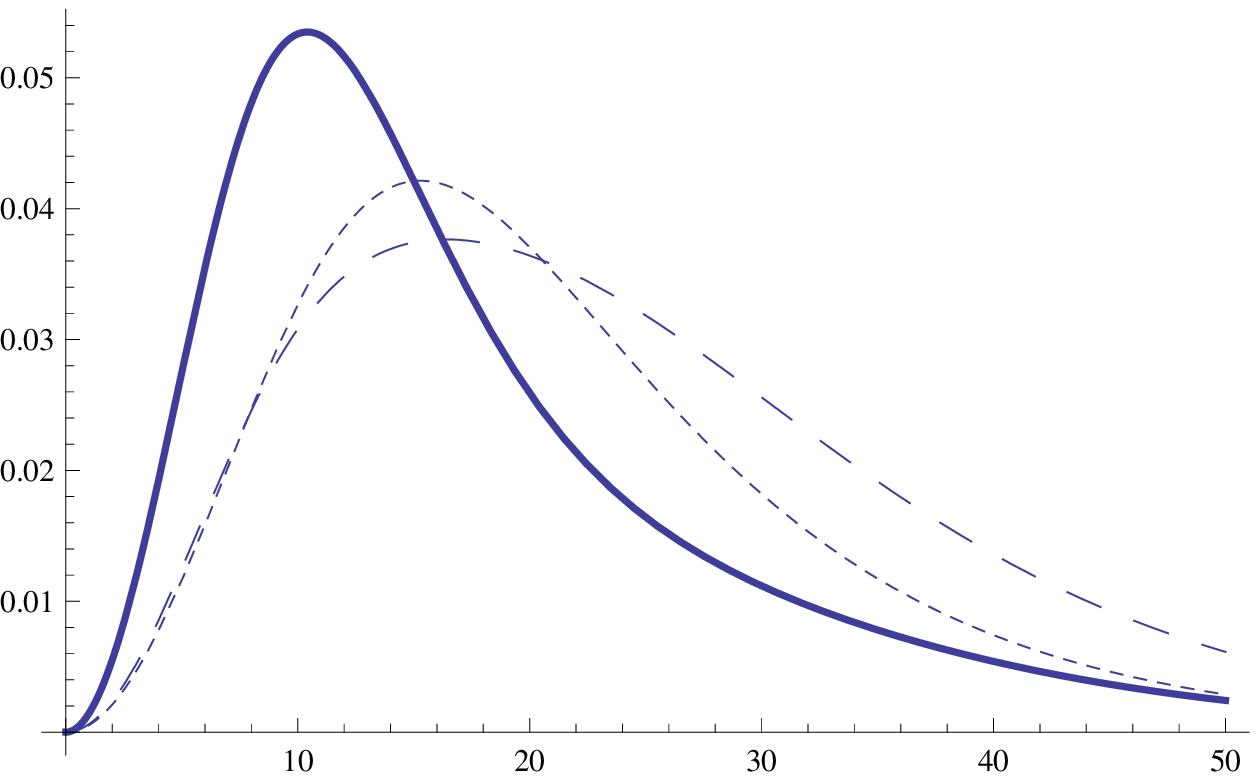}
}
 \subfloat[]
 {
%\rotatebox{90}{\hspace{-0.0cm}{$f_{sp}\rightarrow$MeV$^{-1}$}}
%\rotatebox{90}{\hspace{-0.0cm} {$\Phi(x)\longrightarrow 4 \pi G_Na^2 \rho_0$}}
\rotatebox{90}{\hspace{-0.0cm}{$f_{sp}\rightarrow$MeV$^{-1}$ IH}}
\includegraphics[scale=0.6]{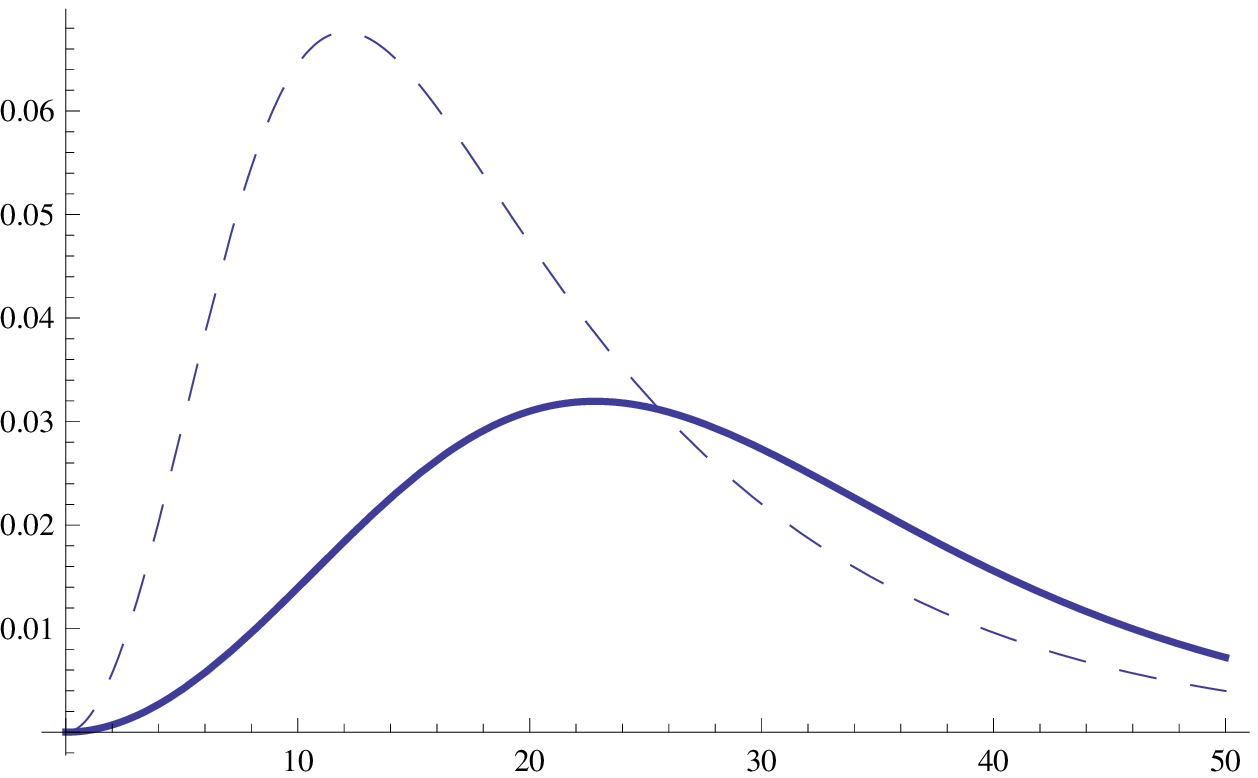}
}
\\
{\hspace{-0.0cm} {$E_{\nu}\rightarrow $MeV}}\\
%{\hspace{-0.0cm} (a) \hspace{4.0cm} (b)}
 \caption{The normalized to unity SN spectrum for $a=3$ (a) and the modified SN spectra also for  $a=3$ in the normal hierarchy scenario  (b) and the inverted hierarchy scenario  (c). The continuous, dotted and dashed curves correspond to $T$=3.5 ($\nu_e$), 5 ($\tilde{\nu}_e$) and 8 ($\nu_x$) respectively. In the case of the inverted hierarchy scenario the spectra for $\nu_e$ and  $\tilde{\nu}_e$ coincide}
 %\end{center}
 \label{fdis}
  \end{center}
  \end{figure}
  The average neutrino energies obtained from this distribution are shown in Table \ref{tab:avener}. 
  \begin{table}[t]
\caption{ The average supernova neutrino energies as a function of the parameters $a$ and $T$.
\label{tab:avener}
}
\begin{center}
%{\footnotesize
\begin{tabular}{|c|c|c|c|}
\hline
$a$& \multicolumn{3}{c|}{$\prec E_{\nu}\succ$ (MeV)}\\
\hline
&$\nu_e$&$\tilde{\nu}_e$&$\sum_x\nu_x$\\
&T=3.5MeV&T=5MeV&T=8MeV\\
\hline
0& 11.0298 & 15.7569 & 25.211 \\
0.75& 11.4504 & 16.3578 & 26.1724 \\
 1.50&12.0787 & 17.2553 & 27.6085 \\
2.00& 12.6194 & 18.0277 & 28.8443 \\
 3.00&13.9733 & 19.9619 & 31.9391 \\
4.00& 15.6313 & 22.3305 & 35.7288 \\
 5.00&17.5179 & 25.0255 & 40.0408\\
 \hline
\end{tabular}
\end{center}
\end{table}

The number of emitted neutrinos \cite{RAFFELT04} can be obtained from the total emitted energy $U_{\nu}=3\times10^{53}$ erg
\beq
N_{\nu}=\frac{U_{\nu}}{\prec E_{\nu} \succ}.
\eeq
The obtained results are shown in Table \ref{tab:Nnu}
  \begin{table}[t]
\caption{ The number of primary neutrinos emitted in a typical supernova explosion  as a function of the parameters $a$ and $T$ in units of $10^{58}$.
\label{tab:Nnu}
}
\begin{center}
%{\footnotesize
\begin{tabular}{|c|c|c|c|}
\hline
$a$& \multicolumn{3}{c|}{${N_{\nu}}/{10^{58}}$ }\\
\hline
&$\nu_e$&$\tilde{\nu}_e$&$\sum_x\nu_x$\\
&T=3.5 MeV&T=5 MeV&T=8 MeV\\
\hline
0& 0.282969 & 0.198079 & 0.495196 \\
0.75 &0.272575 & 0.190802 & 0.477006 \\
 1.50&0.258397 & 0.180878 & 0.452194 \\
 2.00&0.247326 & 0.173128 & 0.43282 \\
3.00& 0.223361 & 0.156353 & 0.390882 \\
 4.00&0.199669 & 0.139768 & 0.349421 \\
5.00& 0.178167 & 0.124717 & 0.311792\\
 \hline
\end{tabular}
\end{center}
\end{table}
The (time averaged) neutrino flux $\Phi_{\nu}=N_{\nu}/(4 \pi D^2) $ at a distance $D=10$ kpc=$3.1\times 10^{22}$cm is given in Table \ref{tab:Phinu}.
  \begin{table}[t]
\caption{ The (time integrated) neutrino flux, in units of $10^{12}$cm$^{-2}$, at a distance 10 kpc from the source.
\label{tab:Phinu}
}
\begin{center}
%{\footnotesize
\begin{tabular}{|c|c|c|c|}
\hline
%$a$& \multicolumn{3}{c|}{$\frac{N_{\nu}}{10^{58}}$ }\\
$a$& \multicolumn{3}{c|}{${\Phi_{\nu}}/{10^{12}\mbox{cm}^{-2}} $}\\
\hline
&$\nu_e$&$\tilde{\nu}_e$&$\sum_x\nu_x$\\
&T=3.5 MeV&T=5 MeV&T=8 Mev\\
\hline
0& 0.234318 & 0.164023 & 0.410057 \\
 0.75&0.225711 & 0.157997 & 0.394994 \\
1.50& 0.213971 & 0.149779 & 0.374448 \\
2.00& 0.204803 & 0.143362 & 0.358405 \\
 3.00&0.184958 & 0.129471 & 0.323677 \\
 4.00&0.16534 & 0.115738 & 0.289345 \\
 5.00&0.147534 & 0.103274 & 0.258185\\
 \hline
\end{tabular}
\end{center}
\end{table}
\section{Modification of the SRN spectra due to neutrino oscillation oscillation}
Even though the neutral current detector is neutrino flavor bound, the neutrino oscillations modify the SRN spectrum. This modification will affect the expected rates since the different flavors have different temperature and chemical potential.  This modification has recently   been discussed \cite{DigheSmirnov00}, \cite{NMNS15}. Neutrino oscillations imply:
\beq
\frac{dN_{\nu_e}}{dE_{\nu}}=U^2_{e1}\frac{dN_{\nu_1}}{dE_{\nu}}+U^2_{e2}\frac{dN_{\nu_2}}{dE_{\nu}}+U^2_{e3}\frac{dN_{\nu_3}}{dE_{\nu}}
\eeq
\beq
\frac{dN_{\nu_x}}{dE_{\nu}}=U^2_{x1}\frac{dN_{\nu_1}}{dE_{\nu}}+U^2_{x2}\frac{dN_{\nu_2}}{dE_{\nu}}+U^2_{x3}\frac{dN_{\nu_3}}{dE_{\nu}},\,x=\mu,\tau
\eeq
for neutrinos and antineutrinos. The superscript zero refers to the primary  neutrino spectra. It can be shown that for the normal hierarchy $m_1<m_2<m_3$
\beq
\frac{dN_{\nu_1}}{dE_{\nu}}\approx \frac{dN^0_{\nu_e}}{dE_{\nu}},\,\frac{dN_{\nu_2}}{dE_{\nu}}\approx \frac{dN^0_{\nu_x}}{dE_{\nu}},\,\frac{dN_{\nu_3}}{dE_{\nu}}\approx \frac{dN^0_{\nu_x}}{dE_{\nu}} \mbox{ NH },
\eeq
while for the inverted hierarchy (IH) $m_3<m_2<m_1$ these equations become:
\beq
\frac{dN_{\nu_1}}{dE_{\nu}}\approx \frac{dN^0_{\nu_x}}{dE_{\nu}},\,\frac{dN_{\nu_2}}{dE_{\nu}}\approx \frac{dN^0_{\nu_x}}{dE_{\nu}},\,\frac{dN_{\nu_3}}{dE_{\nu}}\approx \frac{dN^0_{\nu_e}}{dE_{\nu}} \mbox{ IH },
\eeq
Combining these equations we get for NI:
\beq
\frac{dN_{\nu_e}}{dE_{\nu}}\approx\frac{2}{3}\frac{dN^0_{\nu_e}}{dE_{\nu}}+\frac{1}{3}\frac{dN^0_{\nu_x}}{dE_{\nu}},\,\frac{dN_{\nu_x}}{dE_{\nu}}\approx\frac{1}{6}\frac{dN^0_{\nu_e}}{dE_{\nu}}+\frac{5}{6}\frac{dN^0_{\nu_x}}{dE_{\nu}},
\eeq
while for the IH we find:
\beq
\frac{dN_{\nu_e}}{dE_{\nu}}\approx\frac{dN^0_{\nu_x}}{dE_{\nu}},\,\frac{dN_{\nu_x}}{dE_{\nu}}\approx\frac{1}{2}\left (\frac{dN^0_{\nu_e}}{dE_{\nu}}+\frac{dN^0_{\nu_x}}{dE_{\nu}}\right ).
\eeq
For completeness we mention that as the neutrinos continue to propagate outwards, they encounter a further modification
by the Mikheyev-Smirnov-Wolfenstein (MSW) effect \cite{Wolfenstein}, \cite{MickSmyr85}. It has been shown \cite{ChiuHuangLai15} that 
\barr
\frac{dN^{''}_{\nu_e}}{dE_{\nu}}&\approx& P_m \frac{dN_{\nu_e}}{dE_{\nu}}+(1-P_m)\frac{dN_{\nu_x}}{dE_{\nu}},\nonumber\\ \frac{dN^{''}_{\nu_x}}{dE_{\nu}}&\approx& (1-P_m) \frac{dN_{\nu_e}}{dE_{\nu}}+P_m \frac{dN_{\nu_x}}{dE_{\nu}}
\earr
The parameter $P_m$ depends not only on the mixing angles but also on  the crossing probabilities  PH and PL for the neutrino eigenstates at higher and lower resonances. It is also different for  neutrinos and  antineutrinos, but we will not elaborate further.
\section{The TPC detector}
It will be very interesting to see whether one can gain detailed information about the supernova neutrino spectrum(SN) and perhaps gain information about the neutrino  hierarchy from a neutral current detector with high sensitivity.

To this end we are proposing to use a gaseous spherical TPC detector dedicated for SN neutrino detection through the neutrino-nucleus coherent process. More specifically  to use a gaseous  spherical TPC detector, dedicated to supernova neutrino  detection, exploiting  the neutrino-nucleus neutron coherent process. The first idea is to employ a a small size spherical TPC detector filled with a high pressure noble gas \cite{Gioma08} \cite{Bourga12}. 
Today the spherical detector is used for dark matter search at LSM (.6 m in diameter 10 bar pressure) underground laboratory and for NEWS experiment a future project at SNOLAB (1.5 m, 10 bar)\cite{GGMDGJBN}. Dark matter detector is focused in light-WIMP search using gas target of light elements as H, He and Ne are more sensitive in in the GeV and sub-GeV range, compared to current experiments using of Xe and Ge. During data taking the threshold was set at 30 eV, a second hint to reach low WIMP mass sensitivity.

 For the SN project we could use a conceptual design of detector based on the existing technology and increase detector diameter and pressure to 50 bar taking into account recent developments. 
A key issue for such high-pressure operation is the use of a sensor ball smaller than the current 6.3 mm in diameter. In recent laboratory investigations we have successfully used 2 mm ball sensor. This is coming closer to our goal of 1 mm ball, which would be the ideal size for reaching stable operation at 50 bar. Using such small sensors will require lower operation voltage and therefore induce low electric field (E) at large distances: $$E(r_1)= r_2\frac{V_0}{r_1^2} $$
which shows that the electric field at the periphery is proportional to the radius of the small ball ($r_2$) and inversely proportional to the radius square of the external sphere ($r_1$). Such low field may become a concern for large detector. 
Clearly there is a contradiction between large detector and small sensor. 

Recently we have successfully tested, however, a new idea with a multi-ball sensor which is supposed to solve this problem:
The multi-ball system is employing many small conductive balls arranged around a larger spherical surface, that could also tune the electric field at large distance and at the same time could produce a segmentation of the detector. In the multi-ball arrangement the electric field at large distance is proportional to the distance of the balls from the center of the spherical detector and not to the radius of the balls as it is in the case of existing central single ball detector. Notice that such multi-ball system will optimize the electric field for any size of the external sphere and the detector segmentation will help to avoid pile-up events. 

An enhancement of the neutral current component is achieved via the coherent effect of all neutrons in the target. Thus employing, e.g., Xe at 10 Atm, with a feasible threshold energy of about 100 eV in the recoiling nuclei, we show that one may expect, depending on the neutrino hierarchy, between 300 and 500 events for a sphere of radius 3m. This can go up to 1500 and 2500 events, if the pressure is raised at 50 Atm, something quite feasible to-day taking into account recent detector progress.

\section{The differential and total cross section}
\begin{figure}[!ht]
 \begin{center}
 \subfloat[]
 {
 \rotatebox{90}{\hspace{1.0cm} {$F^2(T_A) \rightarrow $}}
\includegraphics[scale=0.6]{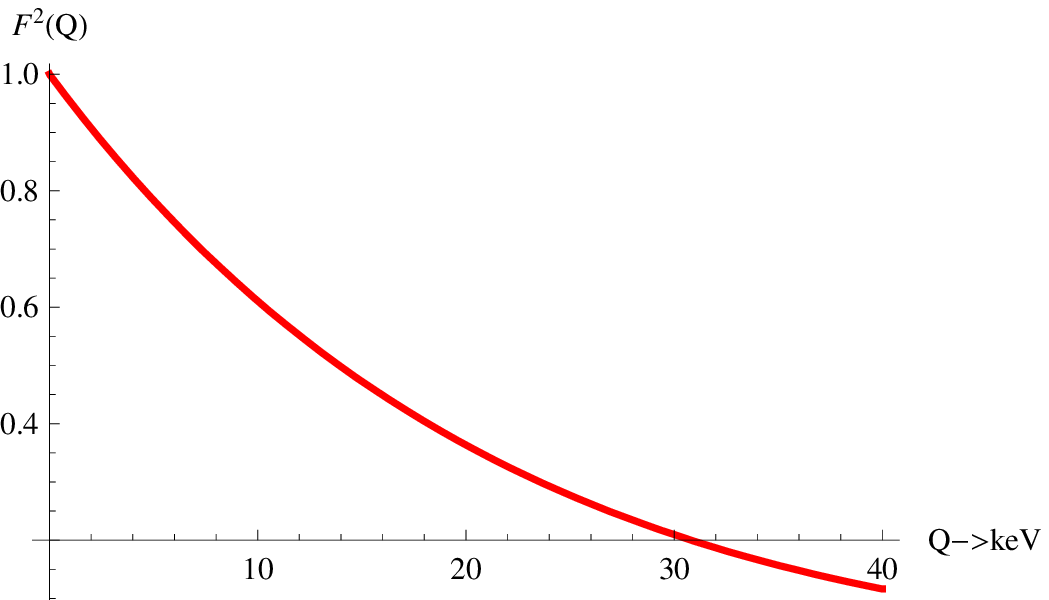}
}
\hspace{8.0cm}$T_A \rightarrow$ keV\\
\subfloat[]
{
 %\rotatebox{90}{\hspace{1.0cm} {$F^2(T_A) \rightarrow $}}
\includegraphics[scale=0.6]{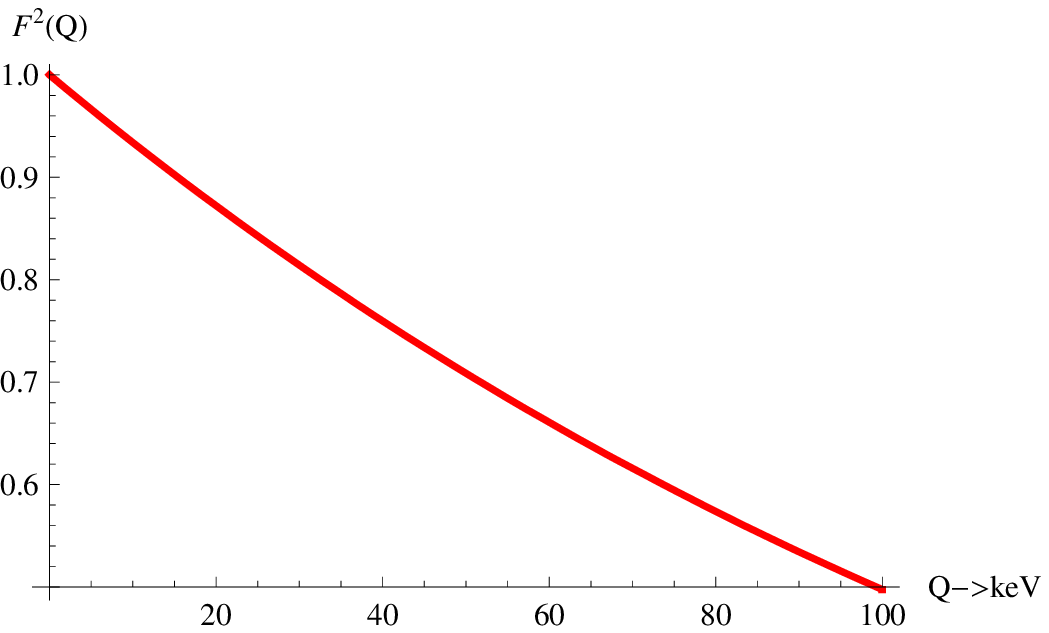}
}\\
\hspace{4.0cm}$T_A \rightarrow$ keV
 \caption{The square of the nuclear form factor, $F^2(T_A)$, as a function of the recoil energy for A=131 (a)
and A=40 (b). Note that the maximum recoil energy is different for each target.}
 \label{fig:ff}
 \end{center}
  \end{figure}

The differential cross  section for a given neutrino energy $E_{\nu}$ can be cast in the form\cite{VERGIOM06}:
\beq
 \left(\frac{d\sigma}{dT_A}\right)_{w}(T_A,E_{\nu})=\frac{G^2_F Am_N}{2 \pi}~(N^2/4) F_{coh}(T_A,E_{\nu}),
%\nonumber\\
 \label{elaswAV1}
\eeq
with
\beq
F_{coh}(T_A,E_{\nu})= F^2(q^2)
  \left ( 1+(1-\frac{T_A}{E_{\nu}})^2
-\frac{Am_NT_A}{E^2_{\nu}} \right)
 \label{elaswAV2}
  \eeq
  where $N$ is the neutron number and $F(q^2)= F(T_A^2+2 A m_N T_A)$ is the nuclear form factor.
  The effect of the nuclear form factor depends on the target (see Fig. \ref{fig:ff}).
  
   Since the SN   source is not "`monochromatic"' the above equation can be written as:
     \beq \frac{d\sigma}{dT_A}=\int_{E(T_A)}^{(E_{\nu})_{\mbox{\tiny{max}}}}\left(\frac{d\sigma}{dT_A}\right)_{w}(T_A,E_{\nu})f_{sp}(E_{\nu},T,a)d E_{\nu}
  \label{elaswAV3}
  \eeq
  Where $(E_{\nu})_{\mbox{\tiny{max}}}$ is the maximum neutrino energy and 
  $$ E(T_A)=\frac{T_A}{2}+\sqrt{\frac{T_A}{2}(M_A+\frac{T_A}{2})}$$
    Here $(E_{\nu})_{\mbox{\tiny{max}}}=\infty$.\\
    Integrating the total cross section of Fig. \ref{Fig:disigma.131} from  $T_A=E_{th}$ to infinity we obtain the total cross section.  The threshold energy $E_{th}$ depends on the detector.
   
  The number of the observed events for each neutrino species is found to be:
  \beq
  N_{ev}(a,T)=\Phi_{\nu} (a,T) \sigma({a,T,E_{th}}) N_{N}(P,T_0,R)
  \label{Tevent}
  \eeq
  \beq
   N_N(P,T_0,R)=\frac{P}{kT_0} \frac{4}{3} \pi R^3
  \eeq
  where $N_N$ is the number of nuclei in the target, which depends on the pressure, ($P$), the absolute temperature, ($T_0$) and the radius $R$ of the detector. We find:
  \beq
  N_N(P,T_0,R)=1.04\times10^{30} \frac{P}{10~\mbox{Atm}} \frac{300\,^0\mbox{K}}{T_0}\left (\frac{R}{10\mbox{m}} \right )^3
  \eeq
      \subsection{Results for the Xe target}
\begin{figure}[!ht]
 \begin{center}
 \subfloat[]
{
\rotatebox{90}{\hspace{-0.0cm}{$\frac{d \sigma}{d T_A}\rightarrow 10^{-39}$}cm$^2/$keV}
%\rotatebox{90}{\hspace{-0.0cm} {$\Phi(x)\longrightarrow 4 \pi G_Na^2 \rho_0$}}
\includegraphics[scale=0.7]{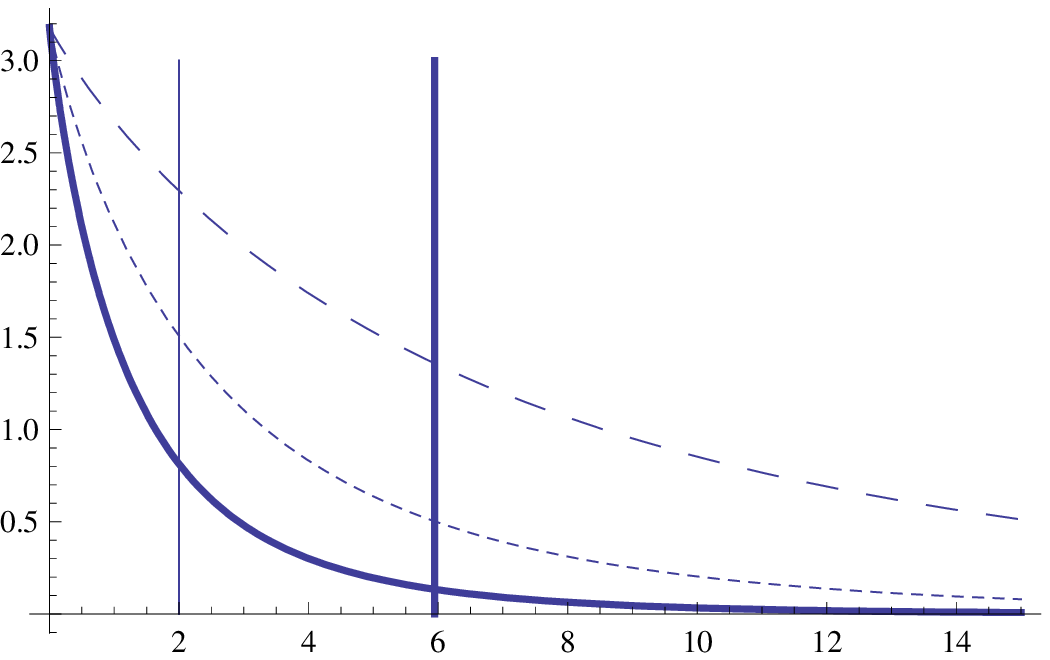}
}\\
%{\hspace{-2.0cm} {$\frac{\Phi(x)}{\Phi_0}\longrightarrow $}}
 \subfloat[]
 {
%\rotatebox{90}{\hspace{-0.0cm}{$f_{sp}\rightarrow$MeV$^{-1}$}}
%\rotatebox{90}{\hspace{-0.0cm} {$\Phi(x)\longrightarrow 4 \pi G_Na^2 \rho_0$}}
\rotatebox{90}{\hspace{-0.0cm}{$\frac{d \sigma}{d T_A}\rightarrow 10^{-39}$}cm$^2/$keV}
\includegraphics[scale=0.7]{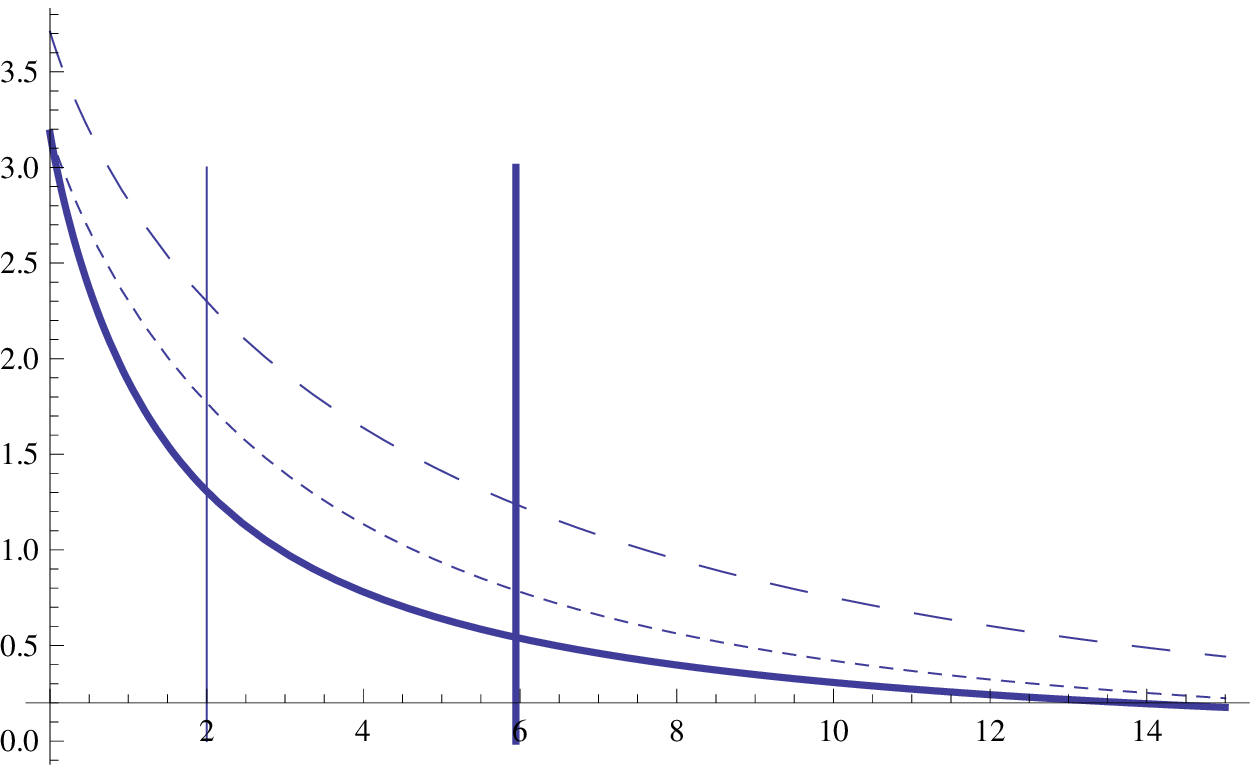}
}
\\
%{\hspace{-2.0cm} {$\frac{\Phi(x)}{\Phi_0}\longrightarrow $}}
 \subfloat[]
 {
%\rotatebox{90}{\hspace{-0.0cm}{$f_{sp}\rightarrow$MeV$^{-1}$}}
%\rotatebox{90}{\hspace{-0.0cm} {$\Phi(x)\longrightarrow 4 \pi G_Na^2 \rho_0$}}
\rotatebox{90}{\hspace{-0.0cm}{$\frac{d \sigma}{d T_A}\rightarrow 10^{-39}$}cm$^2/$keV}
\includegraphics[scale=0.7]{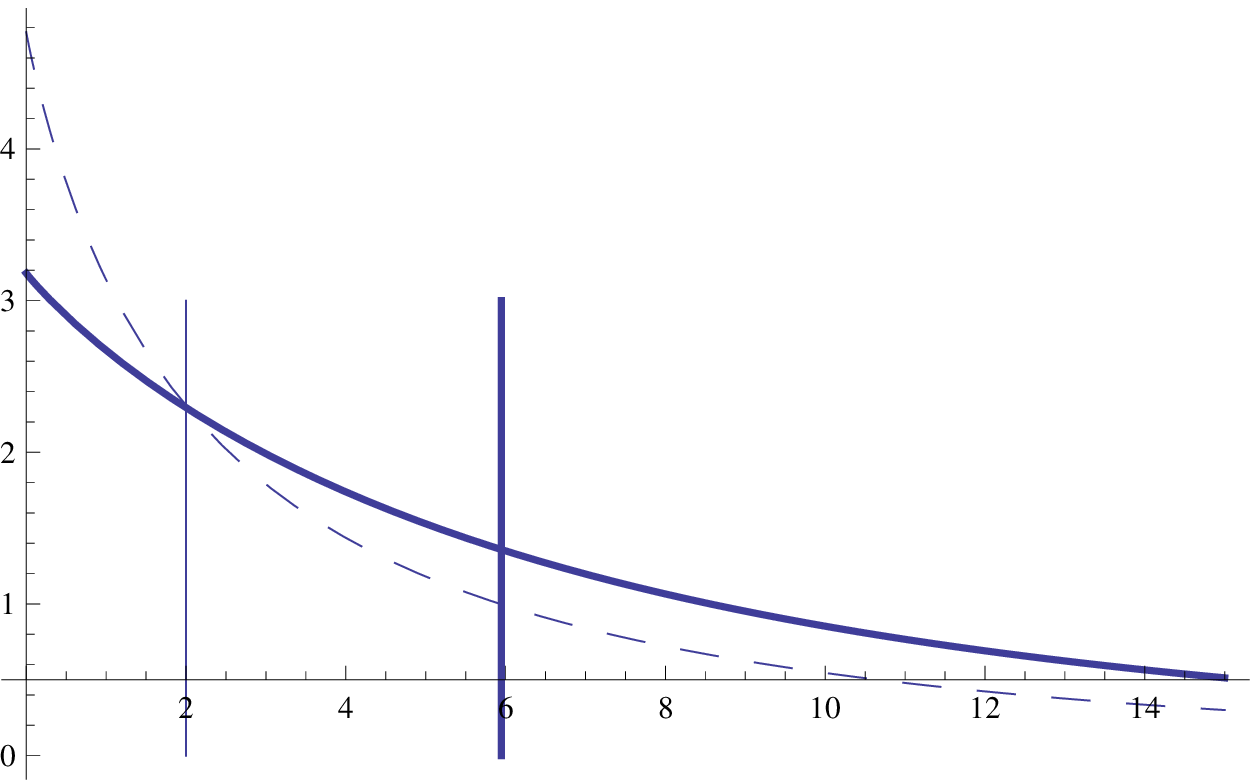}
}
\\
{\hspace{-0.0cm} {$T_{A}\rightarrow $keV}}\\
%{\hspace{-0.0cm} (a) \hspace{4.0cm} (b)}
 \caption{The differential cross section for elastic neutrino nucleus scattering in the case of the target $^{131}_{54}$Xe as a function of the recoil energy $T_A$ in keV. In the case of a threshold energy of 2 keV, only the space on the right of the vertical bar is available. In the presence of quenching  only the space on the right of the thick vertical bar is available. From top to bottom the primordial spectrum (a), the normal  hierarchy (NH) (b) and inverted hierarchy (IH) (c) scenario. Otherwise the notation is the same as in Fig. \ref{fdis}}
 \label{Fig:disigma.131}
  \end{center}
  \end{figure}

\begin{figure}[!ht]
 \begin{center}
 \subfloat[]
{
\rotatebox{90}{\hspace{-0.0cm}{$\frac{\sigma(E_{th})}{\sigma(0)}\rightarrow$}}
%\rotatebox{90}{\hspace{-0.0cm} {$\Phi(x)\longrightarrow 4 \pi G_Na^2 \rho_0$}}
\includegraphics[scale=0.7]{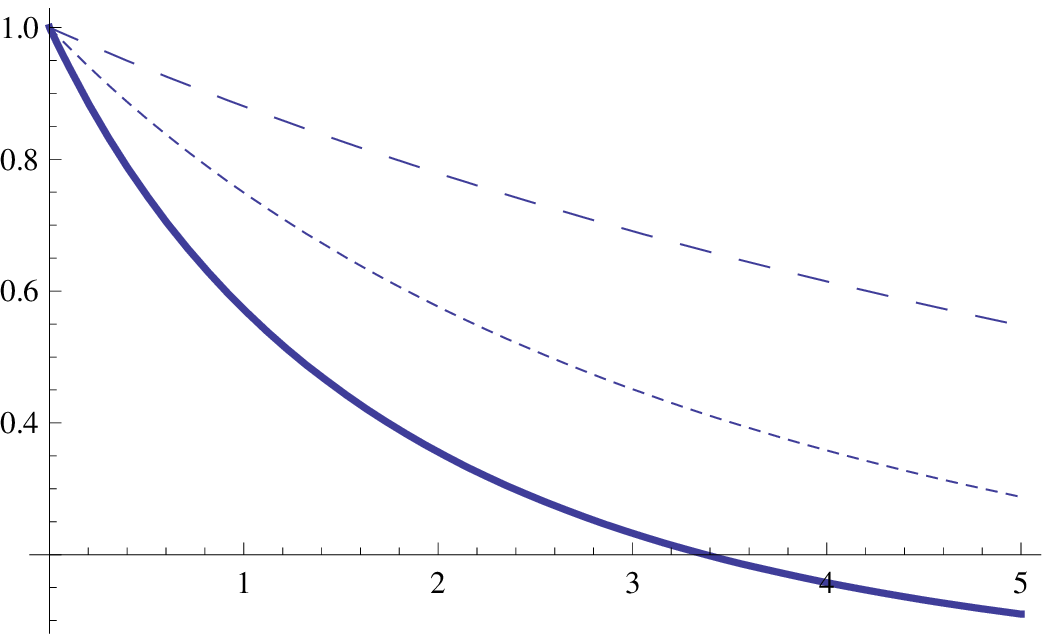}
}\\
%{\hspace{-2.0cm} {$\frac{\Phi(x)}{\Phi_0}\longrightarrow $}}
 \subfloat[]
 {
%\rotatebox{90}{\hspace{-0.0cm}{$f_{sp}\rightarrow$MeV$^{-1}$}}
%\rotatebox{90}{\hspace{-0.0cm} {$\Phi(x)\longrightarrow 4 \pi G_Na^2 \rho_0$}}
\rotatebox{90}{\hspace{-0.0cm}{$\frac{\sigma(E_{th})}{\sigma(0)}\rightarrow$}}
\includegraphics[scale=0.7]{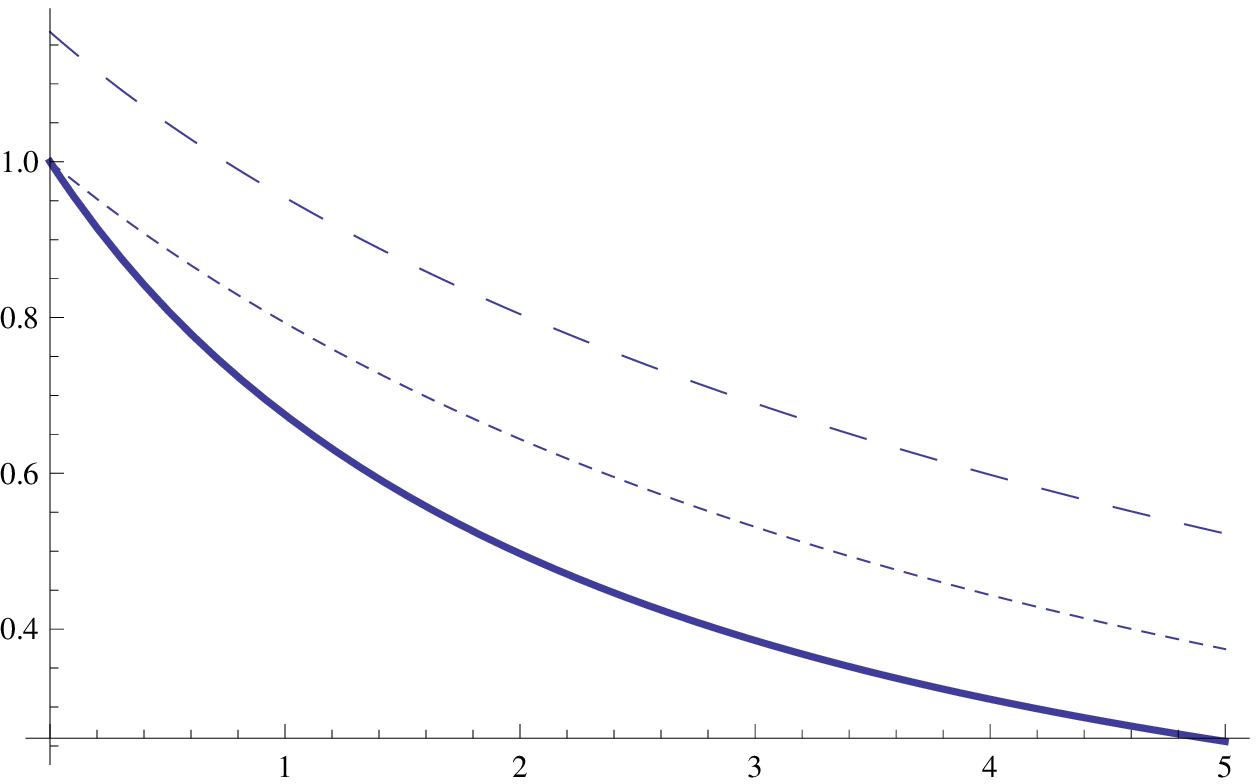}
}
\\
%{\hspace{-2.0cm} {$\frac{\Phi(x)}{\Phi_0}\longrightarrow $}}
 \subfloat[]
 {
%\rotatebox{90}{\hspace{-0.0cm}{$f_{sp}\rightarrow$MeV$^{-1}$}}
%\rotatebox{90}{\hspace{-0.0cm} {$\Phi(x)\longrightarrow 4 \pi G_Na^2 \rho_0$}}
\rotatebox{90}{\hspace{-0.0cm}{$\frac{\sigma(E_{th})}{\sigma(0)}\rightarrow$}}
\includegraphics[scale=0.7]{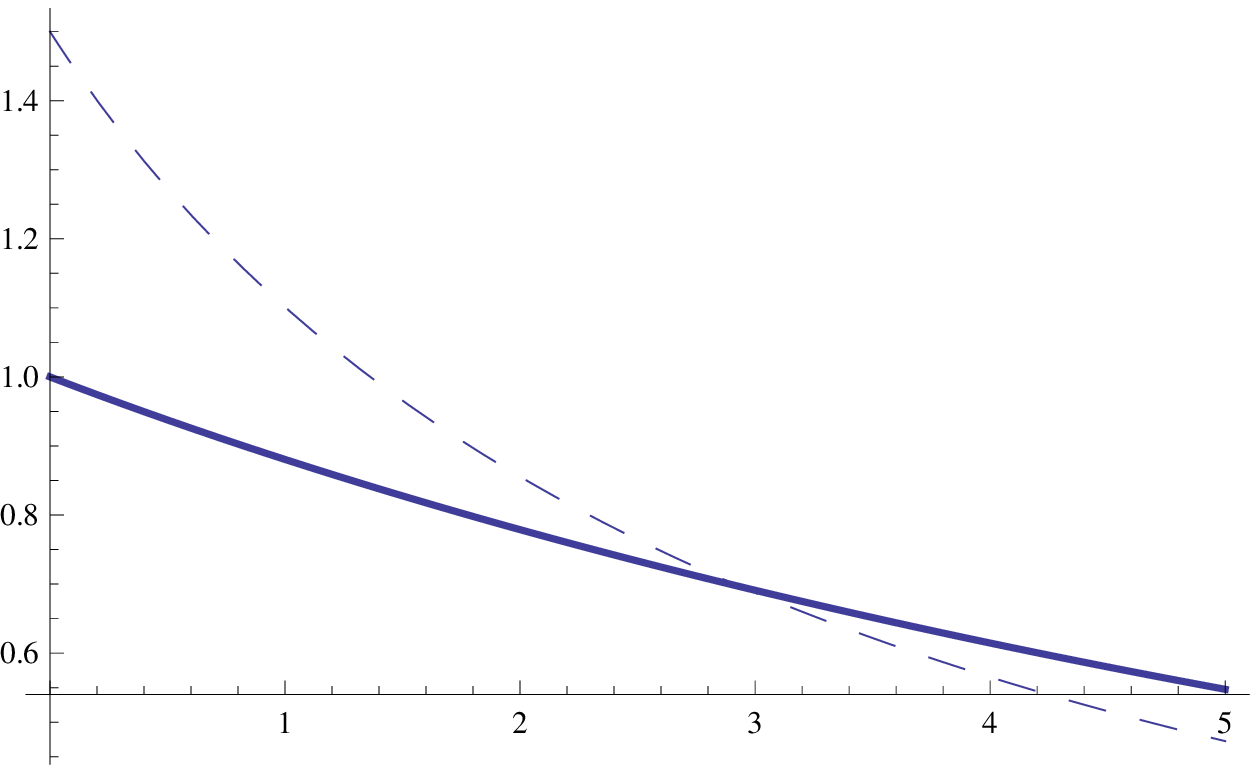}
}
\\
{\hspace{-0.0cm} {$E_{th}\rightarrow $keV}}\\
%{\hspace{-0.0cm} (a) \hspace{4.0cm} (b)}
 \caption{The ratio of the cross section at threshold $E_{th}$ divided by that at zero threshold as a function of the threshold energy in keV in the case of a Xe target. From top to bottom the primordial spectrum (a), the normal  hierarchy (NH) (b) and inverted hierarchy (IH) (c) scenario.
  Otherwise the notation is the same as in Fig. \ref{fdis}}
 \label{Fig:tsigma.131}
  \end{center}
  \end{figure}

The differential cross section for neutrino elastic scattering, obtained with the above neutrino spectrum, on the target $^{131}_{54}$Xe is shown in Fig. \ref{Fig:disigma.131}.
  Integrating the total cross section of Fig. \ref{Fig:disigma.131} from  $T_A=0$ to infinity we obtain the total cross section given in table \ref{tab:sigma.131}.
   \begin{table}[t]
\caption{ The total neutrino nucleus cross section in the case of Xe target in units of $10^{-39}$cm$^{2}$ assuming zero detector threshold. 
\label{tab:sigma.131}
}
\begin{center}
%{\footnotesize
\begin{tabular}{|c|c|c|c|c|}
\hline
%$a$& \multicolumn{3}{c|}{$\frac{N}{10^{58}}$ }\\
$a$& \multicolumn{4}{c|}{${\sigma}/{10^{-39}\mbox{cm}^{2}} $}\\
\hline
&$\nu_e$&$\tilde{\nu}_e$&$\sum_x\nu_x$&Total\\
&T=3.5 MeV&T=5 MeV&T=8 MeV&\\
\hline
 0 & 4.117 & 8.312 & 19.764&32.194 \\
 0.75 & 4.361 & 8.815 & 20.921&34.097 \\
 1.50 & 4.749 & 9.608 & 22.727& 37.083\\
 2.00 & 5.104 & 10.330 & 24.346&39.780 \\
 3.00 & 6.074 & 12.288 & 28.621&46.083 \\
 4.00 & 7.408 & 14.966 & 34.147&56.521 \\
 5.00 & 9.118 & 18.364& 40.546&68.028\\
 \hline
\end{tabular}
\end{center}
\end{table}
The above results refer to an ideal detector operating down to zero energy threshold. In the case of non zero threshold the event rate is suppressed as shown in Fig.  \ref{Fig:tsigma.131}. 
	
       Using Eq. \ref{Tevent} and the above total cross sections, after summing over all neutrino species (i.e. over all T),  we obtain the number of events shown in Table \ref{tab:rate.131}. It is a surprise for us that the total rate increases somewhat  after the modification of the primary neutrino spectra. It is not shown her, bu the e  fraction of $\nu_x$ is is $77\%$, $65\%$ and $46\%$
for the S, NH and IH neutrino spectra respectively, i.e. it deceases as we go from the primary to  the modified  neutrino spectra. This is not relevant for a neutral current detector, but it may be an important element for detectors using the charged current current process, so long as  they are sensitive only to the $\nu_e$ and $\bar{\nu}_e$ components.  

\begin{table}[t]
\caption{ The total event rate as a function of $a$ in the case of a gaseous Xe target under a temperature 300 $^0$K and  various pressures with the indicated spherical detector  radii. S , NH and IH  stand for neutrino spectra in the standard (primary ), modified in the NH and the IH  neutrino respectively. These results were obtained by summing  over all neutrino species assuming a zero detector energy threshold. 
\label{tab:rate.131}
}
\begin{center}
\begin{tabular}{|r|r|r|r|r|r|r|r|r|r|}
\hline
%$a$& \multicolumn{3}{c|}{$\frac{N}{10^{58}}$ }\\
%$a$& \multicolumn{4}{c|}{${\sigma}/{10^{-39}\mbox{cm}^{2}} $}\\
%\hline
a&R=10m&R=10m&R=10m&R=3m&R=3m&R=3m&R=4m&R=4m&R=4m\\
&P=10Atm&P=10 Atm&P=10Atm&P=50Atm&P=50Atm&P=50Atm&P=10Atm&P=10Atm&P=10Atm\\
&S& NH& IH&S&NH&IH&S&NH& IH\\
\hline
 0 & 10872 & 12275 & 15083 & 1467 & 1657 &
   2036 & 695 & 785 & 965 \\
 0.75 & 11089 & 12520 & 15383 & 1497 & 1690
   & 2076 & 709 & 801 & 984 \\
 1.5 & 11427 & 12901 & 15850 & 1542 & 1741
   & 2139 & 731 & 825 & 1014 \\
 2. & 11726 & 13238 & 16262 & 1583 & 1787 &
   2195 & 750 & 847 & 1040 \\
 3. & 12482 & 14089 & 17302 & 1685 & 1902 &
   2335 & 798 & 901 & 1107 \\
 4. & 13378 & 15093 & 18523 & 1806 & 2037 &
   2500 & 856 & 965 & 1185 \\
 5. & 14287 & 16108 & 19749 & 1928 & 2174 &
   2666 & 914 & 1030 & 1263 \\
 \hline
\end{tabular}
\end{center}
\end{table}

In the presence of a detector threshold of even 1 keV the above rates are reduced by about 20$\%$ (50$\%$) in the absence (presence) of quenching.
\subsection{The Ar target}
\begin{figure}[!ht]
 \begin{center}
 \subfloat[]
{
\rotatebox{90}{\hspace{-0.0cm}{$\frac{d \sigma}{d T_A}\rightarrow 10^{-40}$}cm$^2/$keV}
%\rotatebox{90}{\hspace{-0.0cm} {$\Phi(x)\longrightarrow 4 \pi G_Na^2 \rho_0$}}
\includegraphics[scale=0.7]{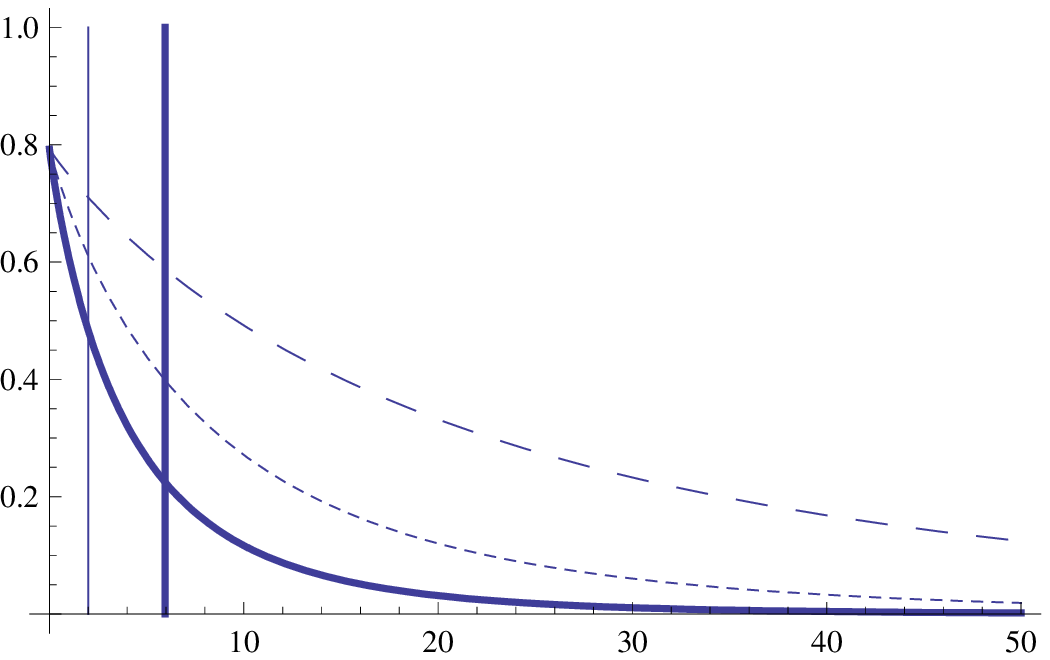}
}\\
%{\hspace{-2.0cm} {$\frac{\Phi(x)}{\Phi_0}\longrightarrow $}}
 \subfloat[]
 {
%\rotatebox{90}{\hspace{-0.0cm}{$f_{sp}\rightarrow$MeV$^{-1}$}}
%\rotatebox{90}{\hspace{-0.0cm} {$\Phi(x)\longrightarrow 4 \pi G_Na^2 \rho_0$}}
\rotatebox{90}{\hspace{-0.0cm}{$\frac{d \sigma}{d T_A}\rightarrow 10^{-40}$}cm$^2/$keV}
\includegraphics[scale=0.7]{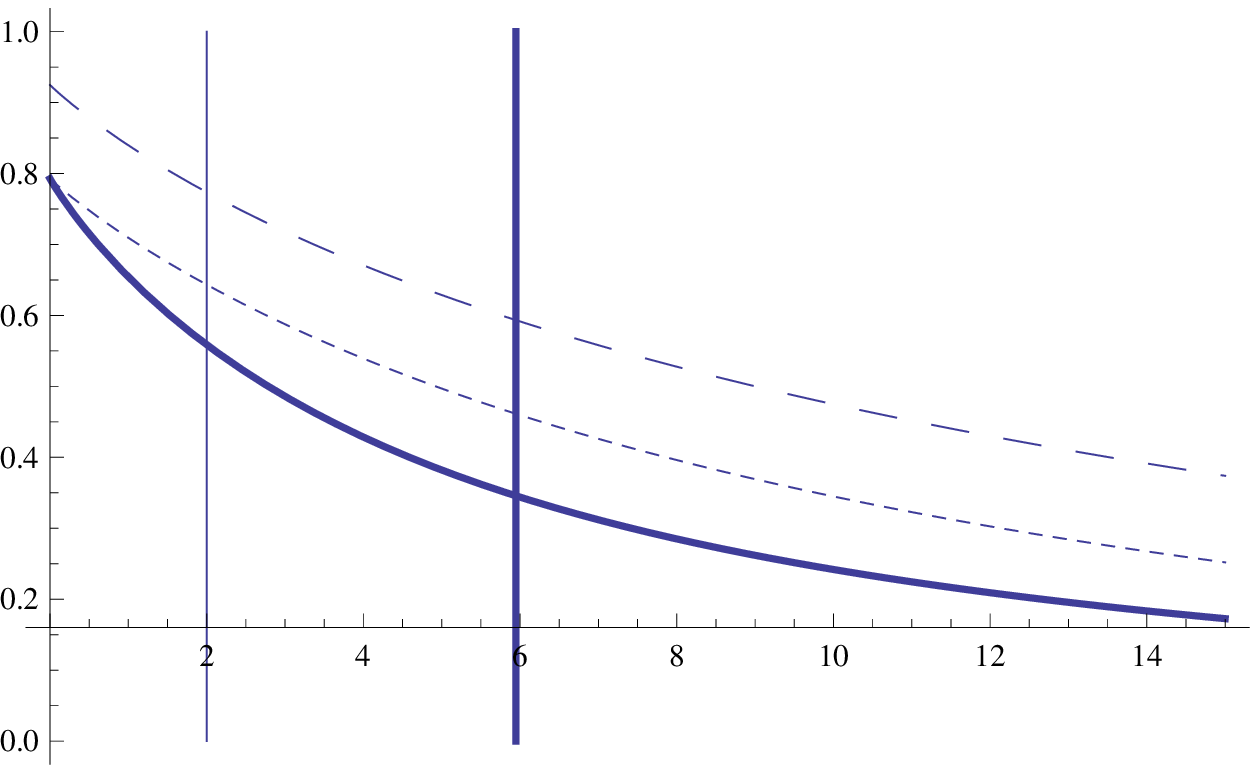}
}
\\
%{\hspace{-2.0cm} {$\frac{\Phi(x)}{\Phi_0}\longrightarrow $}}
 \subfloat[]
 {
%\rotatebox{90}{\hspace{-0.0cm}{$f_{sp}\rightarrow$MeV$^{-1}$}}
%\rotatebox{90}{\hspace{-0.0cm} {$\Phi(x)\longrightarrow 4 \pi G_Na^2 \rho_0$}}
\rotatebox{90}{\hspace{-0.0cm}{$\frac{d \sigma}{d T_A}\rightarrow 10^{-40}$}cm$^2/$keV}
\includegraphics[scale=0.7]{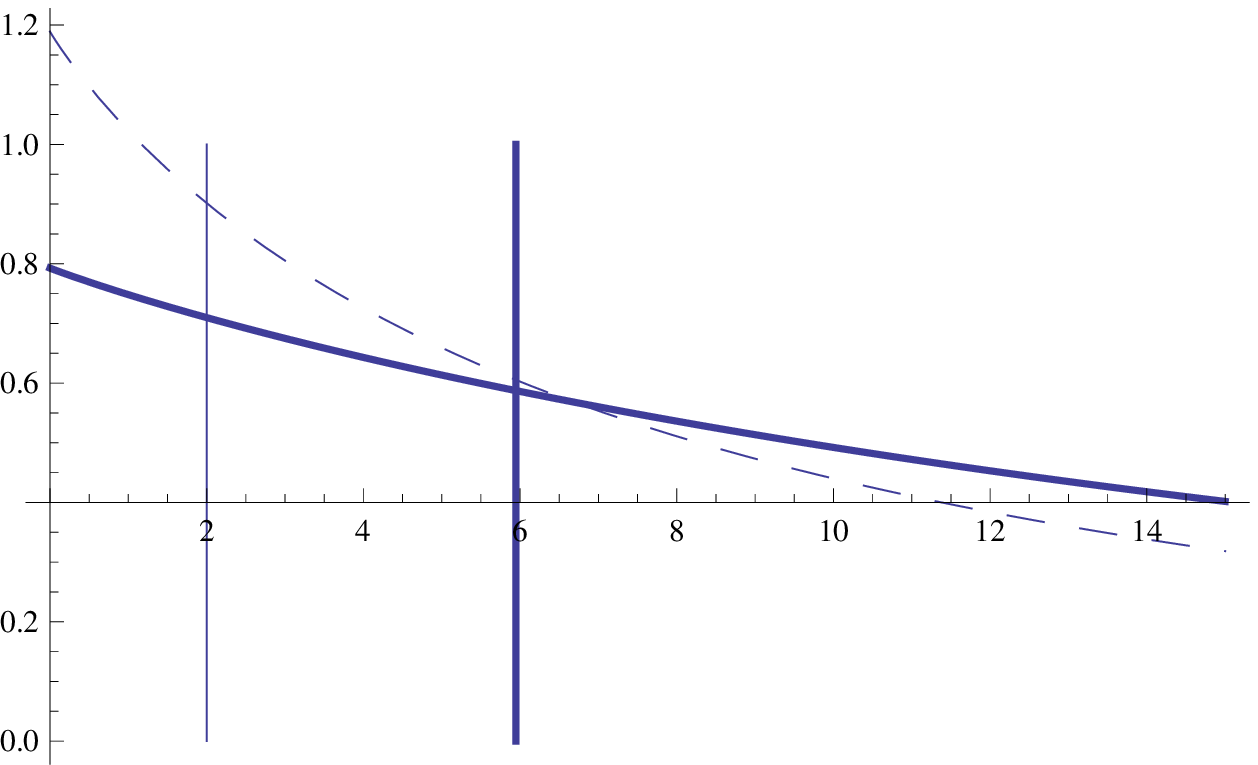}
}
\\
{\hspace{-0.0cm} {$T_{A}\rightarrow $keV}}\\
%{\hspace{-0.0cm} (a) \hspace{4.0cm} (b)}
 \caption{The same as in \ref{Fig:disigma.131} in the case of the Ar target.
 }
 \label{Fig:disigma.40}
  \end{center}
  \end{figure}

The differential cross for neutrino elastic scattering on the target $^{40}_{18}$Ar is shown in Fig. \ref{Fig:disigma.40}. For comparison we are currently calculating the differential cross sections to the excited states of $^{40}$Ar due to the neutral current. We also are going to calculate the charged current cross sections $(\nu_e,e^{-})$ and ($\tilde{\nu}_e,e^{+})$ on $^{40}$Ar, which are of interest in the proposal GLACIER, one of the large detectors\footnote{V. Tsakstara, private communication }.

\begin{table}[t]
\caption{ The total neutrino nucleus cross section in the case of Ar target in units of $10^{-40}$cm$^{2}$ assuming zero detector threshold. 
\label{tab:sigma.40}
}
\begin{center}
%{\footnotesize
\begin{tabular}{|c|c|c|c|c|}
\hline
%$a$& \multicolumn{3}{c|}{$\frac{N}{10^{58}}$ }\\
$a$& \multicolumn{4}{c|}{${\sigma}/{10^{-40}\mbox{cm}^{2}} $}\\
\hline
&$\nu_e$&$\tilde{\nu}_e$&$\sum_x\nu_x$&Total\\
&T=3.5 MeV&T=5 MeV&T=8 MeV&\\
\hline
0 & 3.324 & 6.520 & 13.678 &23.521\\
 0.75 & 3.525 & 6.908 & 14.412&24.845 \\
 1.50& 3.843 & 7.518 & 15.528 &26.888\\
 2.00 & 4.133 & 8.067 & 16.497 &28.693\\
 3.00 & 4.917 & 9.537 & 18.905&33.359 \\
 4.00 & 5.990 & 11.488 & 21.690&39.168 \\
 5.00 & 7.353 & 13.843 & 24.480&45.676\\
 \hline
\end{tabular}
\end{center}
\end{table}
 \begin{figure}[!ht]
 \begin{center}
 \subfloat[]
{
\rotatebox{90}{\hspace{-0.0cm}{$\frac{\sigma(E_{th})}{\sigma(0)}\rightarrow$}}
%\rotatebox{90}{\hspace{-0.0cm} {$\Phi(x)\longrightarrow 4 \pi G_Na^2 \rho_0$}}
\includegraphics[scale=0.7]{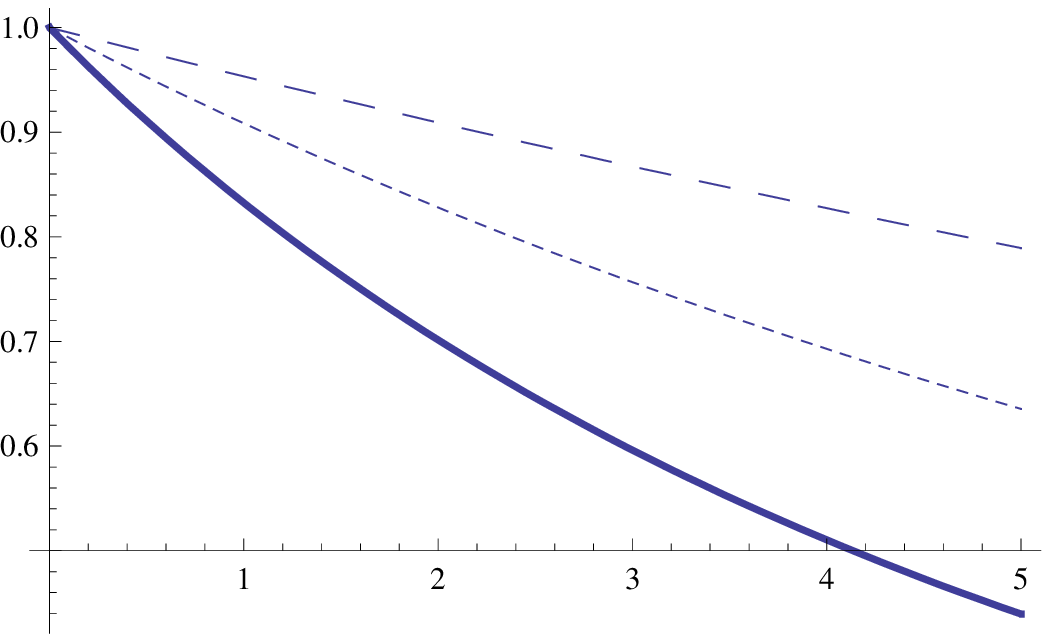}
}\\
%{\hspace{-2.0cm} {$\frac{\Phi(x)}{\Phi_0}\longrightarrow $}}
 \subfloat[]
 {
%\rotatebox{90}{\hspace{-0.0cm}{$f_{sp}\rightarrow$MeV$^{-1}$}}
%\rotatebox{90}{\hspace{-0.0cm} {$\Phi(x)\longrightarrow 4 \pi G_Na^2 \rho_0$}}
\rotatebox{90}{\hspace{-0.0cm}{$\frac{\sigma(E_{th})}{\sigma(0)}\rightarrow$}}
\includegraphics[scale=0.7]{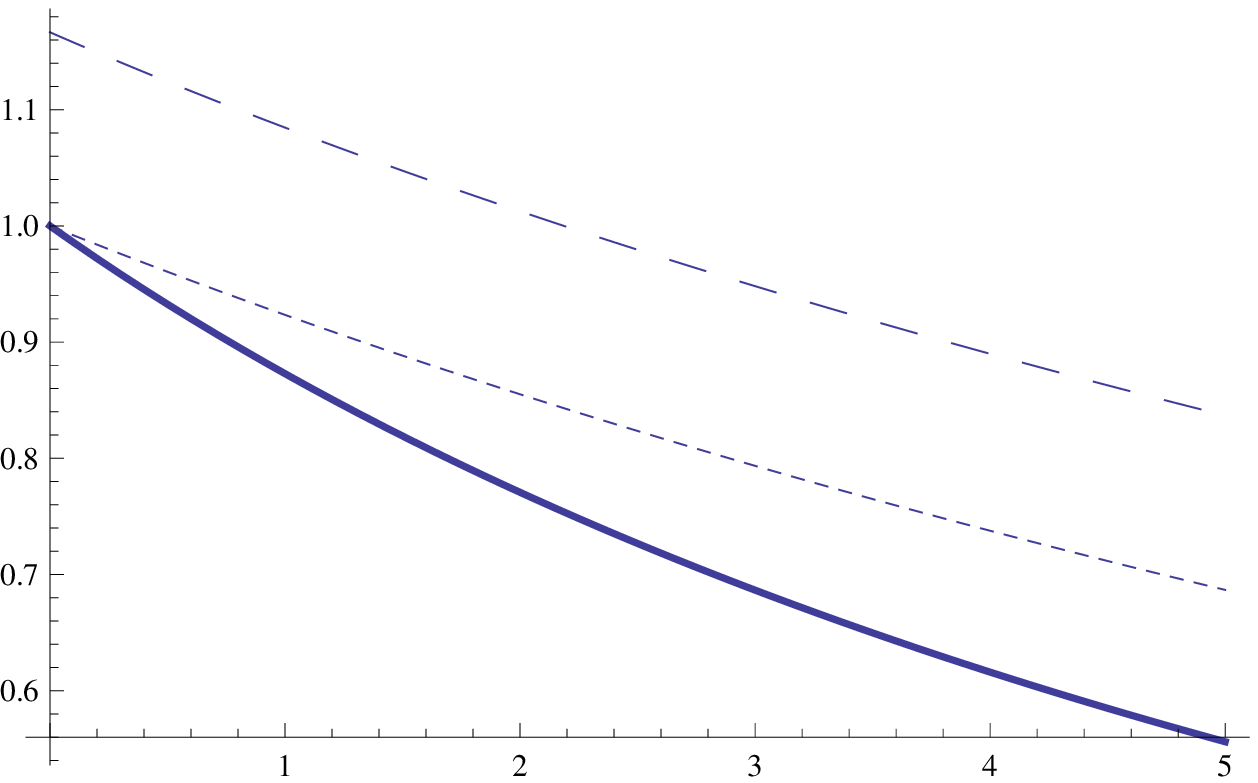}
}
\\
%{\hspace{-2.0cm} {$\frac{\Phi(x)}{\Phi_0}\longrightarrow $}}
 \subfloat[]
 {
%\rotatebox{90}{\hspace{-0.0cm}{$f_{sp}\rightarrow$MeV$^{-1}$}}
%\rotatebox{90}{\hspace{-0.0cm} {$\Phi(x)\longrightarrow 4 \pi G_Na^2 \rho_0$}}
\rotatebox{90}{\hspace{-0.0cm}{$\frac{\sigma(E_{th})}{\sigma(0)}\rightarrow$}}
\includegraphics[scale=0.7]{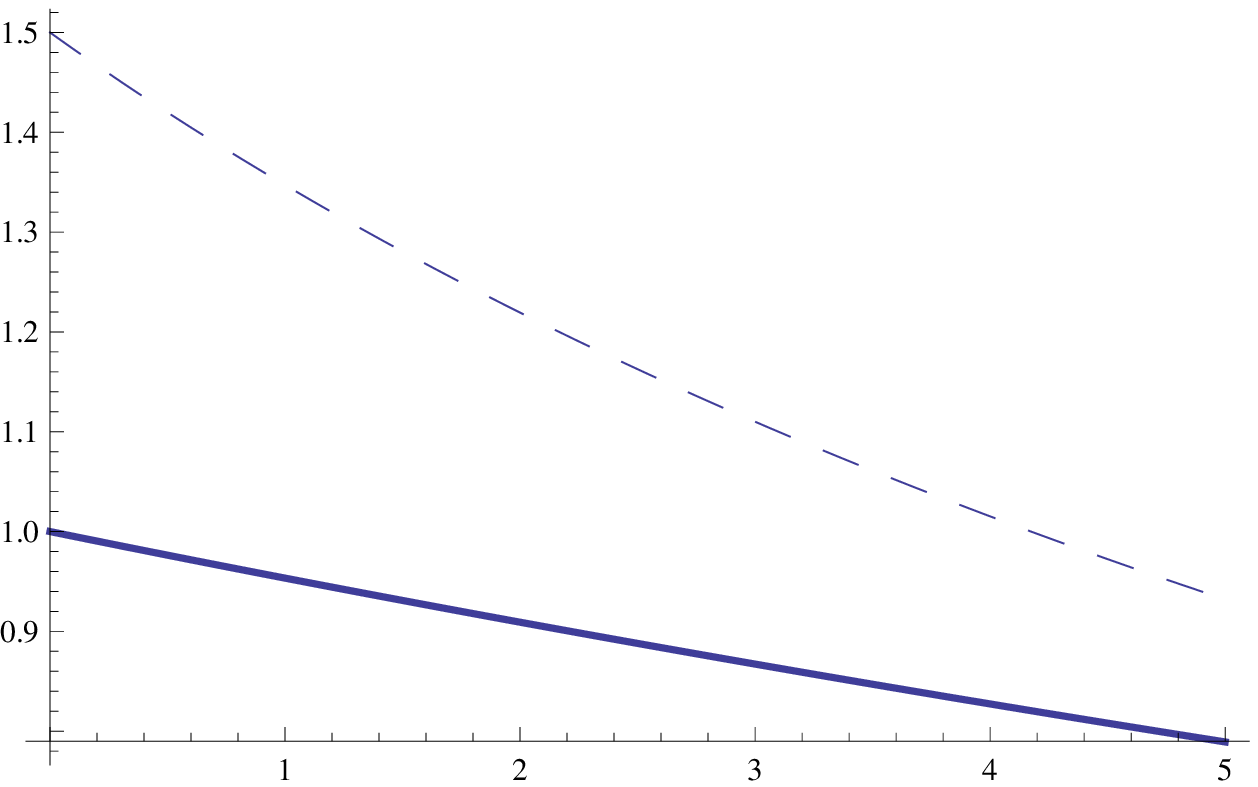}
}
\\
{\hspace{-0.0cm} {$E_{th}\rightarrow $keV}}\\
%{\hspace{-0.0cm} (a) \hspace{4.0cm} (b)}
 \caption{The same as in Fig. \ref{Fig:tsigma.131} in the case of the Ar target}
 \label{Fig:tsigma.40}
  \end{center}
  \end{figure}
 \begin{table}[t]
\caption{ The same as in Table \ref{tab:rate.131} in the case of the target $^{40}$Ar.
\label{tab:rate.40}
}
\begin{center}
\begin{tabular}{|r|r|r|r|r|r|r|r|r|r|}
\hline
%$a$& \multicolumn{3}{c|}{$\frac{N}{10^{58}}$ }\\
%$a$& \multicolumn{4}{c|}{${\sigma}/{10^{-39}\mbox{cm}^{2}} $}\\
%\hline
a&R=10m&R=10m&R=10m&R=3m&R=3m&R=3m&R=4m&R=4m&R=4m\\
&P=10Atm&P=10 Atm&P=10Atm&P=50Atm&P=50Atm&P=50Atm&P=10Atm&P=10Atm&P=10Atm\\
&S& NH& IH&S&NH&IH&S&NH& IH\\
 0 & 777 & 874 & 1070 & 104 & 118 & 144 &
   49 & 55 & 68 \\
 0.75 & 789 & 889 & 1087 & 106 & 120 & 146
   & 50 & 56 & 69 \\
 1.5 & 808 & 910 & 1113 & 109 & 122 & 150 &
   51 & 58 & 71 \\
 2. & 824 & 928 & 1134 & 111 & 125 & 153 &
   52 & 59 & 72 \\
 3. & 861 & 968 & 1182 & 116 & 130 & 159 &
   55 & 61 & 75 \\
 4. & 895 & 1005 & 1225 & 120 & 135 & 165 &
   57 & 64 & 78 \\
 5. & 920 & 1031 & 1254 & 124 & 139 & 169 &
   58 & 66 & 80 \\
 \hline
\end{tabular}
\end{center}
\end{table}
%  \begin{table}[t]
%\caption{ The same as in Table \ref{tab:rate.131} in the case of the target Ar. }
%\label{tab:rate.40}
%\begin{center}
%%{\footnotesize
%\begin{tabular}{|r|r|r|}
%\hline
%$a$& \multicolumn{3}{c|}{$\frac{N}{10^{58}}$ }\\
%$a$& \multicolumn{4}{c|}{${\sigma}/{10^{-39}\mbox{cm}^{2}} $}\\
%\hline
%a&R=10m&R=4m\\
%\hline
% 0 & 193 & 12 \\
% 0.75 & 197 & 13 \\
% 1.50 & 203 & 13 \\
% 2.00 & 209 & 13 \\
% 3.00 & 223 & 14 \\
% 4.00 & 242 & 15 \\
% 5.00 & 262 & 17\\
%  \hline
%\end{tabular}
%\end{center}
%\end{table}
In the presence of a detector threshold of even 1 keV the above rates are reduced by about 10$\%$ (30$\%$) in the absence (presence) of quenching.
\subsection{The Ne target}
\begin{figure}[!ht]
 \begin{center}
 \subfloat[]
{
\rotatebox{90}{\hspace{-0.0cm}{$\frac{d \sigma}{d T_A}\rightarrow 10^{-41}$}cm$^2/$keV}
%\rotatebox{90}{\hspace{-0.0cm} {$\Phi(x)\longrightarrow 4 \pi G_Na^2 \rho_0$}}
\includegraphics[scale=0.7]{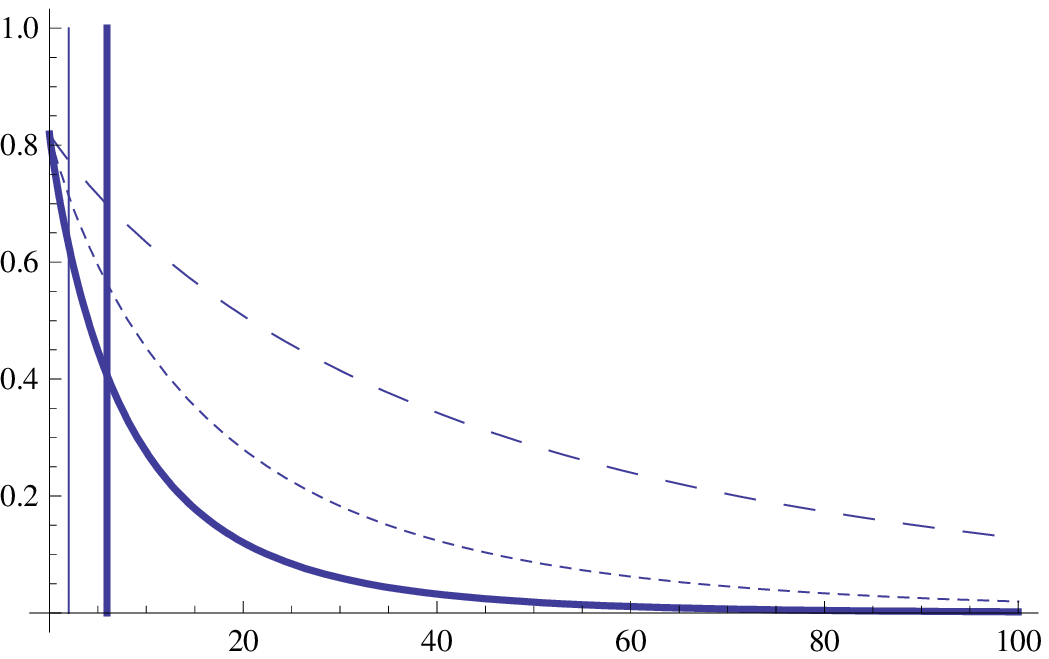}
}\\
%{\hspace{-2.0cm} {$\frac{\Phi(x)}{\Phi_0}\longrightarrow $}}
 \subfloat[]
 {
%\rotatebox{90}{\hspace{-0.0cm}{$f_{sp}\rightarrow$MeV$^{-1}$}}
%\rotatebox{90}{\hspace{-0.0cm} {$\Phi(x)\longrightarrow 4 \pi G_Na^2 \rho_0$}}
\rotatebox{90}{\hspace{-0.0cm}{$\frac{d \sigma}{d T_A}\rightarrow 10^{-41}$}cm$^2/$keV}
\includegraphics[scale=0.7]{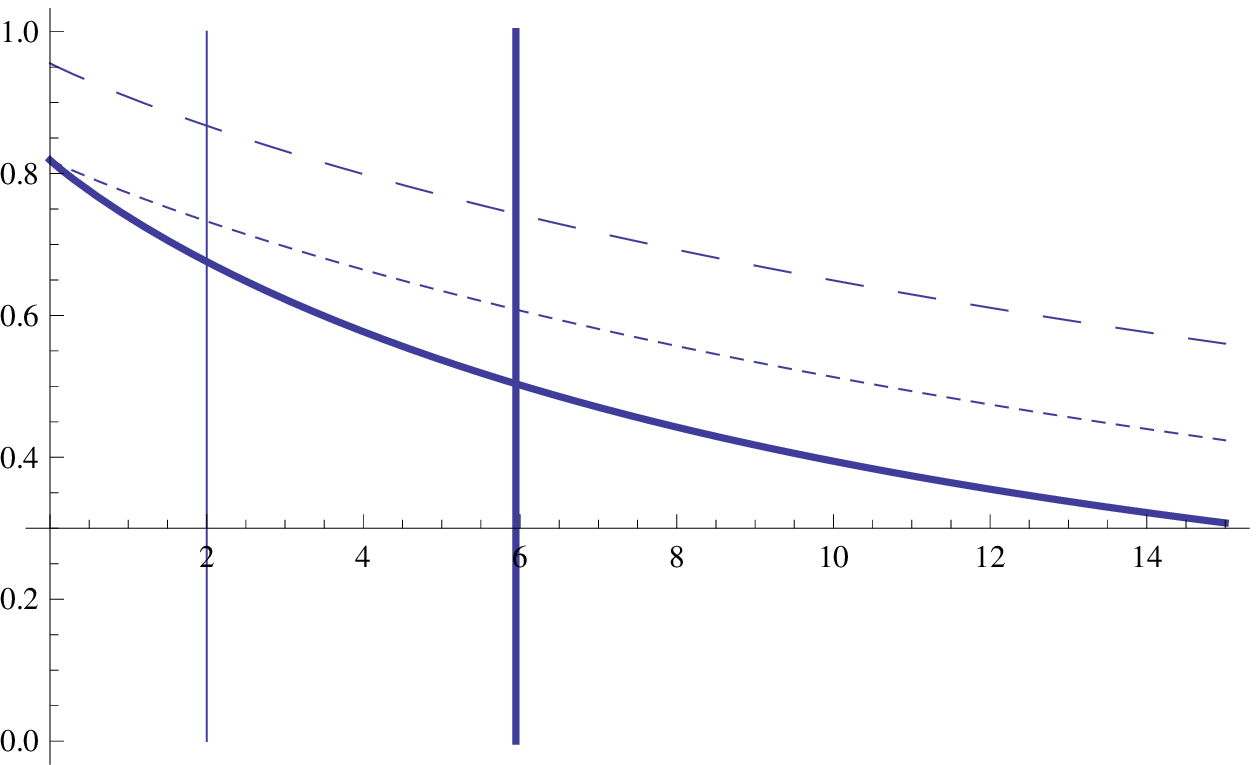}
}
\\
%{\hspace{-2.0cm} {$\frac{\Phi(x)}{\Phi_0}\longrightarrow $}}
 \subfloat[]
 {
%\rotatebox{90}{\hspace{-0.0cm}{$f_{sp}\rightarrow$MeV$^{-1}$}}
%\rotatebox{90}{\hspace{-0.0cm} {$\Phi(x)\longrightarrow 4 \pi G_Na^2 \rho_0$}}
\rotatebox{90}{\hspace{-0.0cm}{$\frac{d \sigma}{d T_A}\rightarrow 10^{-41}$}cm$^2/$keV}
\includegraphics[scale=0.7]{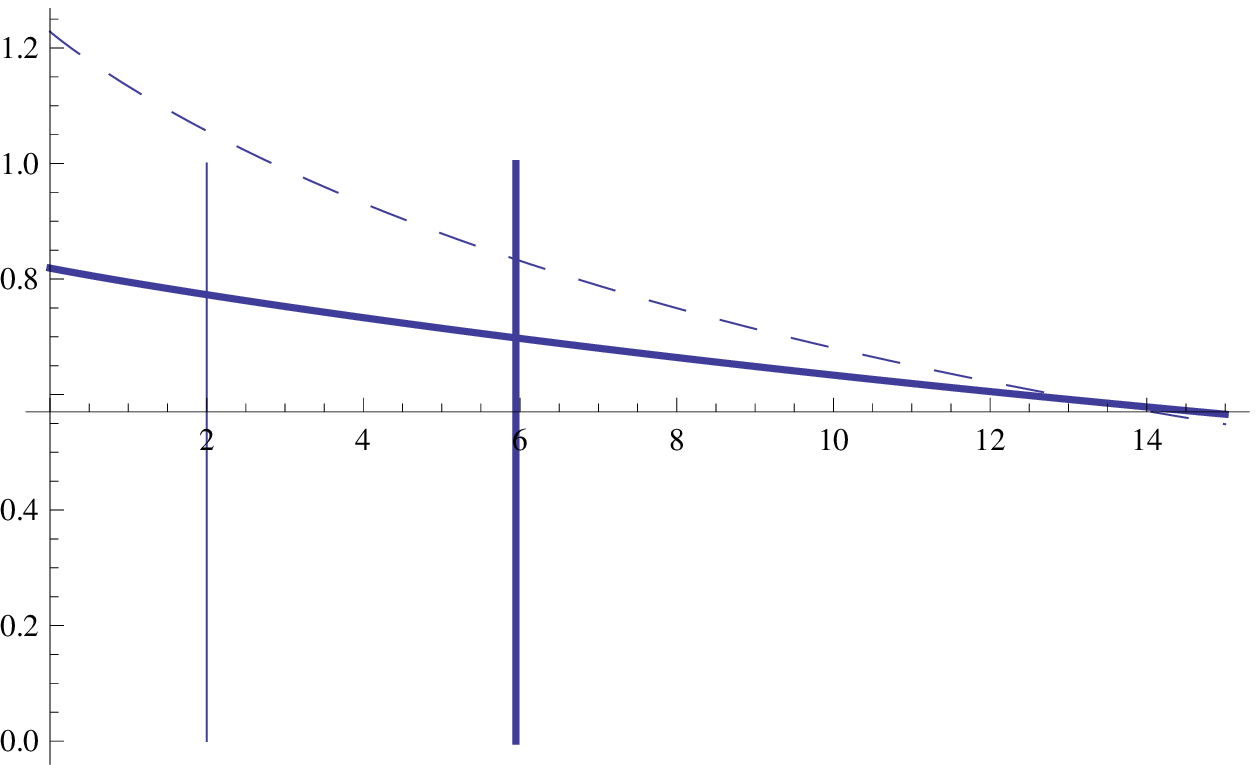}
}
\\
{\hspace{-0.0cm} {$T_{A}\rightarrow $keV}}\\
%{\hspace{-0.0cm} (a) \hspace{4.0cm} (b)}
 \caption{The same as in \ref{Fig:disigma.131} in the case of the Ne target.
 }
 \label{Fig:disigma.20}
  \end{center}
  \end{figure}

The differential cross for neutrino elastic scattering on the target $^{20}_{10}$Ne is shown in Fig. \ref{Fig:disigma.20}.
   
   \begin{table}[t]
\caption{ The total neutrino nucleus cross section in the case of Ne target in units of $10^{-41}$cm$^{2}$ assuming zero detector threshold. 
\label{tab:sigma.20}
}
\begin{center}
%{\footnotesize
\begin{tabular}{|c|c|c|c|c|}
\hline
%$a$& \multicolumn{3}{c|}{$\frac{N}{10^{58}}$ }\\
$a$& \multicolumn{4}{c|}{${\sigma}/{10^{-41}\mbox{cm}^{2}} $}\\
\hline
&$\nu_e$&$\tilde{\nu}_e$&$\sum_x\nu_x$&Total\\
&T=3.5 MeV&T=5 MeV&T=8 MeV&\\
\hline
 0 & 6.861 & 13.456 & 28.232&48.548 \\
 0.75 & 7.277 & 14.258 & 29.747&51.281 \\
 1.50 & 7.934 & 15.515 & 32.049&55.497 \\
 2.00 & 8.531 & 16.649 & 34.049 &59.229\\
 3.00 & 10.150 & 19.683 & 39.021&68.854 \\
 4.00 & 12.364 & 23.708 & 44.772&80.844 \\
 5.00 & 15.176 & 28.568 & 50.537&94.281\\
 \hline
\end{tabular}
\end{center}
\end{table}

    \begin{figure}[!ht]
 \begin{center}
 \subfloat[]
{
\rotatebox{90}{\hspace{-0.0cm}{$\frac{\sigma(E_{th})}{\sigma(0)}\rightarrow$}}
%\rotatebox{90}{\hspace{-0.0cm} {$\Phi(x)\longrightarrow 4 \pi G_Na^2 \rho_0$}}
\includegraphics[scale=0.7]{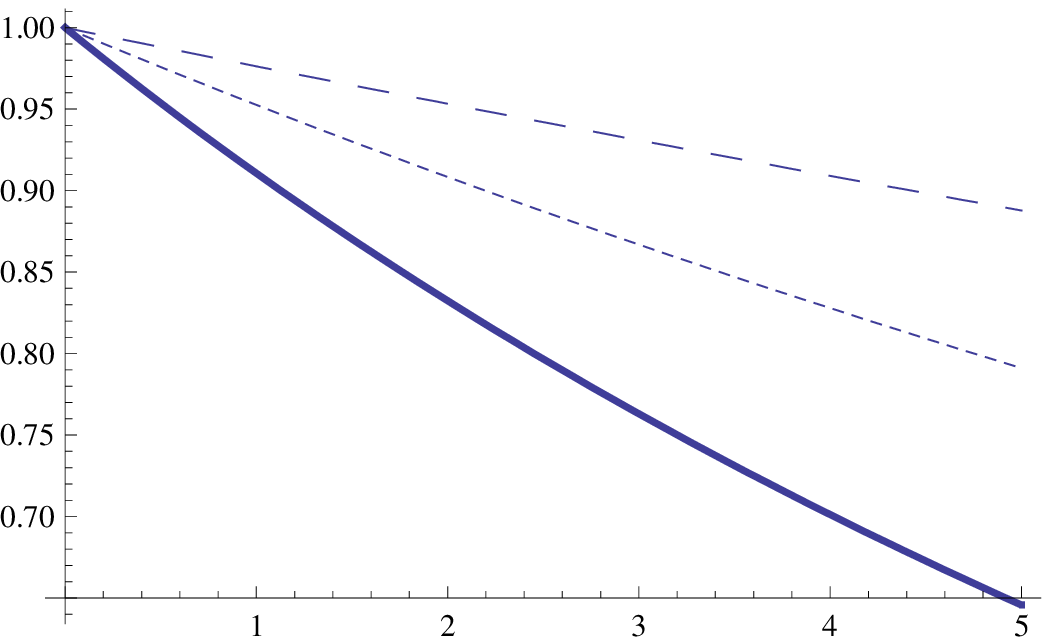}
}\\
%{\hspace{-2.0cm} {$\frac{\Phi(x)}{\Phi_0}\longrightarrow $}}
 \subfloat[]
 {
%\rotatebox{90}{\hspace{-0.0cm}{$f_{sp}\rightarrow$MeV$^{-1}$}}
%\rotatebox{90}{\hspace{-0.0cm} {$\Phi(x)\longrightarrow 4 \pi G_Na^2 \rho_0$}}
\rotatebox{90}{\hspace{-0.0cm}{$\frac{\sigma(E_{th})}{\sigma(0)}\rightarrow$}}
\includegraphics[scale=0.7]{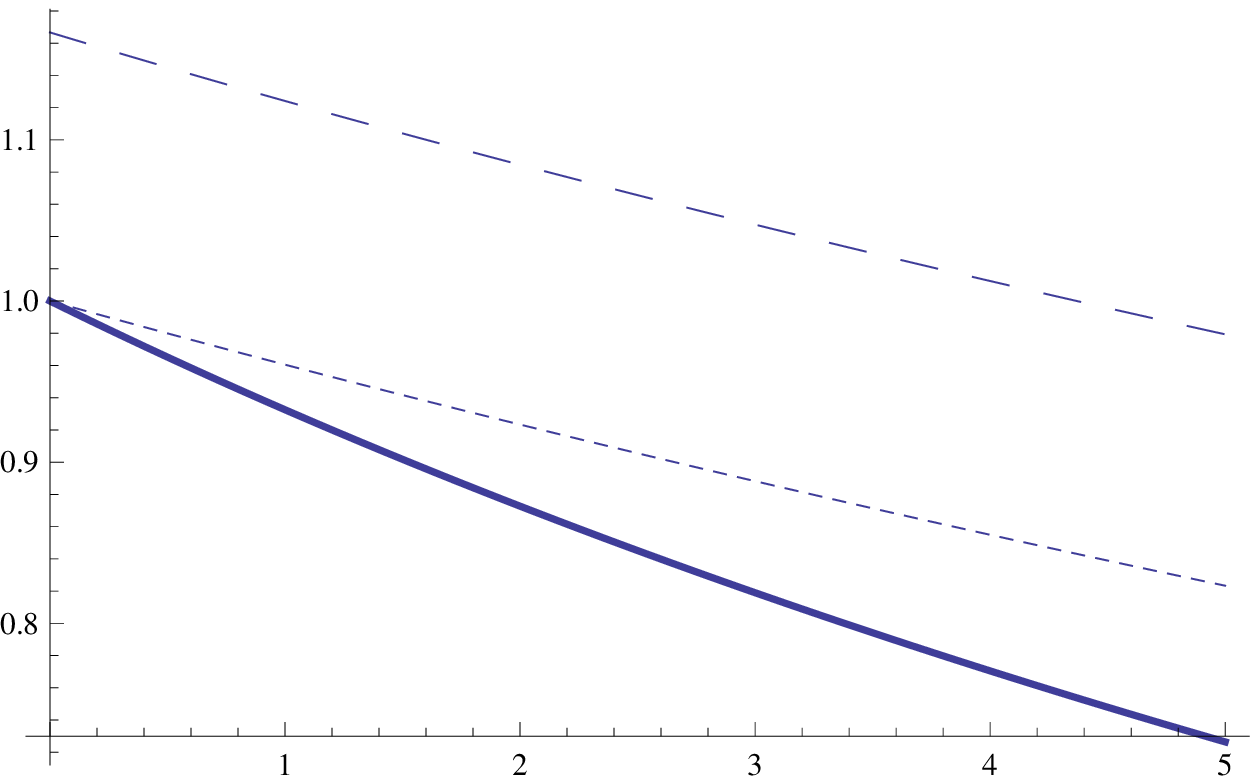}
}
\\
%{\hspace{-2.0cm} {$\frac{\Phi(x)}{\Phi_0}\longrightarrow $}}
 \subfloat[]
 {
%\rotatebox{90}{\hspace{-0.0cm}{$f_{sp}\rightarrow$MeV$^{-1}$}}
%\rotatebox{90}{\hspace{-0.0cm} {$\Phi(x)\longrightarrow 4 \pi G_Na^2 9\rho_0$}}
\rotatebox{90}{\hspace{-0.0cm}{$\frac{\sigma(E_{th})}{\sigma(0)}\rightarrow$}}

\includegraphics[scale=0.7]{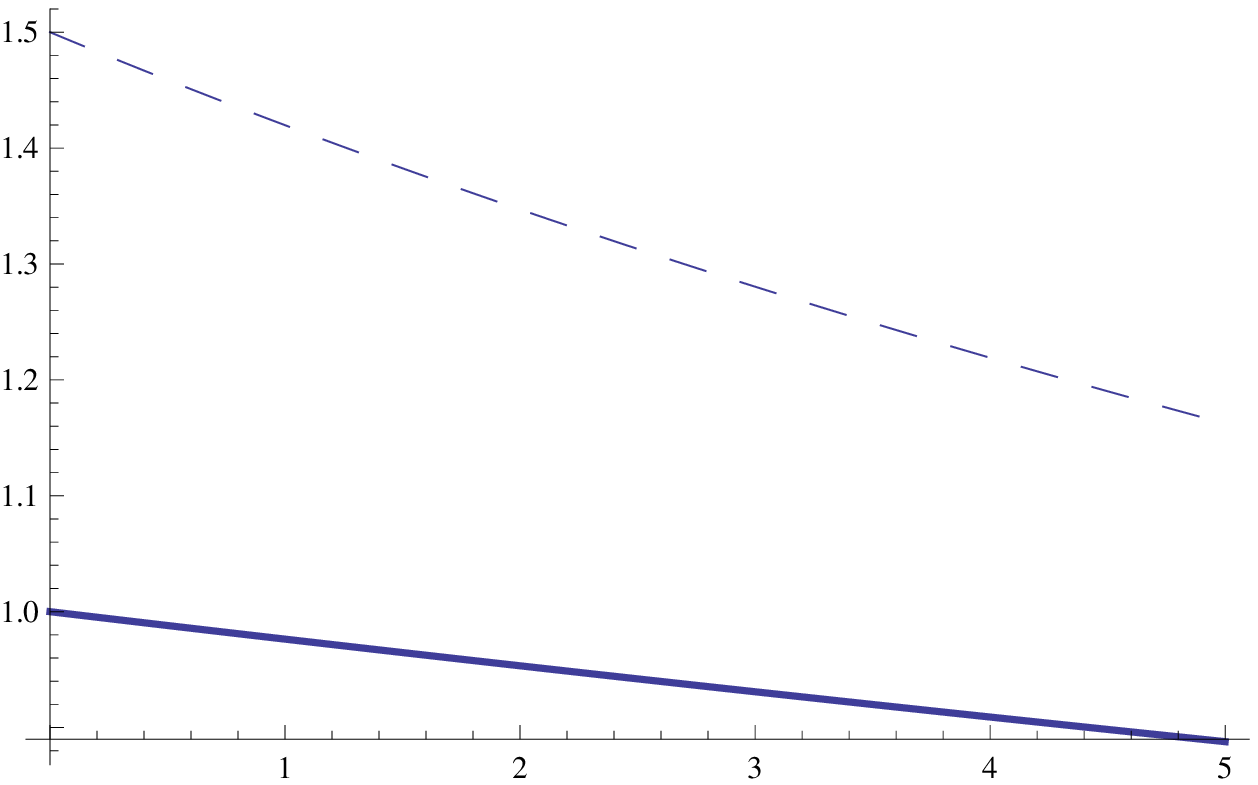}
}
\\
{\hspace{-0.0cm} {$E_{th}\rightarrow $keV}}\\
%{\hspace{-0.0cm} (a) \hspace{4.0cm} (b)}
\caption{The same as in Fig. \ref{Fig:tsigma.131} in the case of the Ne target.}
 \label{Fig:tsigma.20}
  \end{center}
  \end{figure}
  
 \begin{table}[t]
\caption{ The same as in Table \ref{tab:rate.131} in the case of the target $^{20}$Ne.
\label{tab:rate.20}
}
\begin{center}
\begin{tabular}{|r|r|r|r|r|r|r|r|r|r|}
\hline
%$a$& \multicolumn{3}{c|}{$\frac{N}{10^{58}}$ }\\
%$a$& \multicolumn{4}{c|}{${\sigma}/{10^{-39}\mbox{cm}^{2}} $}\\
%\hline
a&R=10m&R=10m&R=10m&R=3m&R=3m&R=3m&R=4m&R=4m&R=4m\\
&P=10Atm&P=10 Atm&P=10Atm&P=50Atm&P=50Atm&P=50Atm&P=10Atm&P=10Atm&P=10Atm\\
&S& NH& IH&S&NH&IH&S&NH& IH\\
 \hline
 0 & 160 & 180 & 220 & 21 & 24 & 29 & 10 &
   11 & 14 \\
 0.75 & 163 & 183 & 224 & 22 & 24 & 30 & 10
   & 11 & 14 \\
 1.5 & 166 & 187 & 229 & 22 & 25 & 31 & 10
   & 12 & 14 \\
 2. & 170 & 191 & 234 & 22 & 25 & 31 & 10 &
   12 & 14 \\
 3. & 177 & 199 & 243 & 23 & 26 & 32 & 11 &
   12 & 15 \\
 4. & 184 & 207 & 253 & 24 & 28 & 34 & 11 &
   13 & 16 \\
 5. & 190 & 213 & 258 & 25 & 28 & 34 & 12 &
   13 & 16 \\
 \hline
\end{tabular}
\end{center}
\end{table}

In the presence of a detector threshold of even  1 keV the above rates are reduced by about 5$\%$ (10$\%$ ) in the absence (presence) of quenching.
\section{Conclusions}
From the above results one can clearly see the advantages of a gaseous spherical TPC detector dedicated for SN neutrino detection. 
The first idea is to employ a small size spherical TPC detector filled with a high
pressure noble gas.  An enhancement of the neutral current component is achieved via the coherent
effect of all neutrons in the target. Thus employing, e.g., Xe at $10$ Atm, with a feasible threshold energy
of about $100$ eV in  the recoiling nuclei,
 one expects, depending on the neutrino hierarchy, between $300$ and $500$ events for a sphere of radius 3m. This can go up to 1500 and 2500 events if the pressure is raised at 50 Atm, something quite feasible even to-day. Employing $^{40}$Ar  one expects between 150 and 200 events but it can become larger if the pressure can be increased above 50 Atm, something quite realistic.
 
The second idea  is to  build several such low cost and robust detectors and install them in 
several places over the world, utilizing the small radius spheres filled with Ar under relatively small pressure. The small low pressure sphees  First estimates show that the required background level
is modest and therefore there is no need for a deep underground laboratory. A mere  100 meter 
water equivalent
coverage seems to be sufficient to reduce the cosmic muon flux at the required level 
(in the case of many such detectors in coincidence, a modest shield is sufficient). 
The maintenance of such systems, quite simple and needed only once every  few years, could be easily assured by universities or even by 
secondary schools, with only 
%Thanks to the simplicity of the system it could be operated by young 
 specific running programs. 
Admittedly  such a detector scheme, measuring low energy nuclear recoils from neutrino 
nucleus elastic scattering, will not be able to  determine the incident neutrino vector and, therefore, 
it is not possible  to localize  the supernova this way. This can be achieved by a cluster of such detectors in 
coincidence by a triangulation technique. \\
 A network of such detectors in coincidence with a sub-keV threshold could also be used to observe
 unexpected low energy events. This low energy range has never been explored using massive
 detectors. A challenge of great importance will be the synchronization of such a detector cluster
with the astronomical $\gamma$-ray burst telescopes to establish whether low energy recoils are
emitted in coincidence with the mysterious $\gamma$ bursts. \\ 
 In summary: networks of  such dedicated gaseous TPC detectors, made out of simple, robust and cheap technology,
 can be simply managed by an international scientific consortium and operated by students. This network
 comprises a system, which can be cheaply maintained
for several decades (or even centuries). Obviously this is   is a key point towards preparing to observe
 few galactic supernova explosions.

 Thus with adequate funding and if we are lucky to soon have   a supernova not much further than 10 kps, we might be able to observe supernova neutrinos. This event, in conjunction with organizing an activity to honor Gerry, will be a tribute to to  G. E. Brown for the memorable work he did on the equation of state of collapsing stars, in collaboration with another giant of physics, H. Bethe. Among other things  they contributed to the understanding of the birth of  our supernova neutrinos.

%For a more complete investcation of the supernova neutrino spectra, one should detect many channels involving %not only neutral but charged currents as well. The expected rates  of these experiments\cite{LLDAPP07} are summarized in table  \ref{tab:LENAGLACIER}
%\begin{table}
%\caption{  Our competitors; The  big beasts
%\label{tab:LENAGLACIER}
%}
%\begin{center}
%{\footnotesize
%\begin{tabular}{|c|c|c|c|c|c|c|}
%\hline
%\hline
%& \multicolumn{2}{c|}{LENA(50\,kton)}& \multicolumn{2}{c|}{GLACIER($\ge$50\,kton)}\\
% \hline
%& reaction&rate&reaction&rate\\
% \hline
% & $\tilde{\nu}_e+p \rightarrow n+e^+$&  $9.0\times 10^{3}$& &\\
% &$\nu_x$\,p\,ES&  $7.0\times 10^{3}$& &\\
% &$\nu_x$\,e\,ES&  $6.0\times 10^{2}$& $\nu_x$\,e\,ES&$1.0\times 10^{3}$\\
% &$\nu_x(^{12}C,^{12}C)$& $3.0\times 10^{3}$& $\nu_x(^{40}Ar,^{40}Ar)$&$3.0\times 10^{4}$ \\
% &$\tilde{\nu}_e(^{12}C,^{12}B)e^{+}$&  $5.0\times 10^{2}$&$\tilde{\nu}_e(^{40}Ar,^{40}Cl)e^{+}$&$5.4\times %10^{2}$ \\
%& $\nu_e(^{12}C,^{12}N)e^{-}$& $8.5\times 10^{1}$ &$\nu_e(^{40}Ar,^{40}K)e^{-}$&$2.0\times 10^{4}$ \\
%\hline
%\hline
%\end{tabular}
%\end{center}
%\end{table}
\section*{Acknowledgement}
JDV would like to acknowledge support by ARC Centre of Excellence in Particle Physics at the Terascale, University of Adelaide, and thank Professor A. W. Thomas for his hospitality. He would also like to express his appreciation to Tom Kuo for providing the opportunity to dedicate this work to the unforgettable  Gerry.    
\section*{References}
%\bibliography{Tenu}

\end{document}